\documentclass[twocolumn]{autart}
\usepackage{lineno}
\usepackage{amsmath,amssymb}
\usepackage{multirow}
\usepackage{graphicx}
\usepackage{color}
\usepackage{natbib}

\newtheorem{assumption}{Assumption}

\newcommand{\beq}{\begin{equation}}
\newcommand{\eeq}{\end{equation}}
\newcommand{\bea}{\begin{eqnarray}}
\newcommand{\eea}{\end{eqnarray}}
\newcommand{\beas}{\begin{eqnarray*}}
\newcommand{\eeas}{\end{eqnarray*}}
\newcommand{\ben}{\begin{enumerate}}
\newcommand{\een}{\end{enumerate}}
\newcommand{\ba}{\begin{array}}

\newcommand{\ea}{\end{array}}

\makeatother
\input tcilatex
\QQQ{Language}{
American English
}

\begin{document}

\begin{frontmatter}

\title{Leading Impulse Response Identification \\ via the Weighted Elastic Net Criterion}



\author{Giuseppe C.~Calafiore}\ead{\\giuseppe.calafiore@polito.it},
\author{Carlo Novara}\ead{\\carlo.novara@polito.it},
\author{Michele Taragna}\ead{\\michele.taragna@polito.it}
\address{Dipartimento di Automatica e Informatica, Politecnico di Torino,\\
Corso Duca degli Abruzzi~24, I--10129, Torino, Italy.}

\begin{abstract}
This paper deals with the problem of finding a low-complexity estimate of
the impulse response of a linear time-invariant discrete-time dynamic system
from noise-corrupted input-output data. To this purpose, we introduce an
identification criterion formed by the average (over the input
perturbations) of a standard prediction error cost, plus a weighted $\ell _1$
regularization term which promotes sparse solutions. While it is well known
that such criteria do provide solutions with many zeros, a critical issue in
our identification context is \emph{where} these zeros are located, since
sensible low-order models should be zero in the tail of the impulse
response. The flavor of the key results in this paper is that, under quite
standard assumptions (such as i.i.d.\ input and noise sequences and system
stability), the estimate of the impulse response resulting from the proposed
criterion is indeed identically zero from a certain time index $n_l$ (named
the \emph{leading order}) onwards, with arbitrarily high probability, for a
sufficiently large data cardinality $N$. Numerical experiments are reported
that support the theoretical results, and comparisons are made with some
other state-of-the-art methodologies.
\end{abstract}


\end{frontmatter}

\section{Introduction}

\label{sec:introduction} A large part of the literature on identification of
linear time-invariant (LTI) dynamic systems follows a statistical approach (%
\cite{Ljung:99_book,SoSt89}), where probabilistic assumptions are made, at
least on the noise corrupting the measurements. The techniques available in
this context may be classified in two main categories: parametric and
nonparametric. Parametric techniques are mainly based on the prediction
error methods (PEMs) or on the maximum likelihood approach, if Gaussian
noise is assumed. The identified models belong to finite-dimensional spaces
of given order, like FIR, ARX, ARMAX, OE, Laguerre, Kautz or orthonormal
basis function models. In order to limit the model complexity and to avoid
possible overfitting, a tradeoff between bias and variance is usually
considered, and the model order selection is performed by optimizing some
suitable cost function -- such as the Akaike's information criterion AIC (%
\cite{Akai74}), the Rissanen's Minimum Description Length MDL, or the
Bayesian information criterion BIC (\cite{Riss78,Schw78}) -- and by applying
some form of cross validation (CV), like hold-out or leave-one-out. Possible
limits of these parametric methods have been pointed out in \cite
{PiDN10,PiCD11,ChOL12}, where it is shown that the sample properties of PEM
approaches equipped with, e.g., AIC\ and CV, may be rather unsatisfactory
and quite far from those predicted by standard (i.e., without model
selection) statistical theory.

The nonparametric techniques aim to obtain the overall system's impulse
response as a suitable deconvolution of observed input-output data. In
particular, very promising approaches have been recently developed, based on
results coming from the machine learning field, see, e.g., \cite{PDCD14} and
the references therein. Rather than postulating finite-dimensional
hypothesis spaces, the estimation problem is tackled in an
infinite-dimensional space, and the intrinsical ill-posedness of the problem
is circumvented by using suitable regularization methods. In particular, the
system's impulse response is modeled as a zero-mean Gaussian process, and
the prior information is introduced by simply assigning a specific
covariance, named \emph{kernel} in the machine learning literature. This
procedure can be interpreted as the counterpart of model order selection in
the parametric PEM approach and, in some cases, it is shown to be much more
robust.

In the present paper, a novel nonparametric method is presented, whereby an
estimate of the system's impulse response is obtained by minimizing a
suitable cost function that directly takes into account the resulting model
complexity. The aim is indeed to obtain a low-complexity model of the
system, in the form of a reduced-order FIR (in this sense, the approach is
not so far from parametric techniques). A key feature of the proposed
approach, representing a relevant improvement over the state of the art, is
that it allows for an effective model order selection, without using strong
a-priori information on the true system. More specifically, we propose the
use of an identification criterion which is a weighted combination of \emph{%
(a)} a standard prediction error term, \emph{(b)} an $\ell _2$
regularization term, and \emph{(c)} a weighted $\ell _1$ penalty term which
promotes sparse solutions; a full justification for such criterion is given
in Section~\ref{sec:elnet}. This type of criterion corresponds to the
so-called Elastic Net cost, which recently became popular in the machine
learning community, see, e.g., \cite{ZouHas:05,DeDeRo:09}. Notice that,
while it is well known that the use of $\ell _1$ regularization leads to
sparse solutions, sparsity alone is not a very interesting feature in our
identification context. Indeed, reduced-order models are obtained only if
the sparsity of the solution follows a specific pattern, whereby the zeros
are all concentrated in the tail of the impulse response. Obtaining such a
pattern is not obvious, nor a-priori granted by the $\ell _1$
regularization. One of the key contributions of this paper is to prove that,
under standard assumptions, the impulse response estimated via our
Elastic-Net type of criterion has the property of being indeed nonzero only
on the initial part of the impulse response (which we shall name the \emph{%
leading response}), with arbitrarily high probability, if the number of data
$N$ is sufficiently large.

The present paper is organized as follows. In Section~\ref{sec:not_prelim}
the notation is set, and some preliminary results on a Chebyshev's type of
convergence for random variables are stated. Section~\ref{sec:setup}
describes the linear identification problem of interest, and contains the
derivations of the Elastic Net cost. The main results on the recovery of the
leading part of the impulse response are contained in Section~\ref{sec:lrr}.
Section~\ref{sec:id_proc} illustrates a practical procedure for implementing
the proposed identification scheme. Numerical experiments, including a
comparative discussion with other identification methods, are given in
Section~\ref{sec:example}. All proofs are contained in the Appendix.

\section{Notation and preliminaries}

\label{sec:not_prelim}

\subsection{Notation}

\label{sec:notation}For a vector $x\in {\ {\mathbb R}^N}$ , we denote by $%
[x]_i$ the $i$-th entry of $x$, and we define its \emph{support} as

\noindent
\[
\mathop{\mathrm{supp}}(x)\doteq \{i\in \{1,\ldots ,N\}:[x]_i\neq 0\}.
\]

\vspace*{-0.3em}

\noindent The notation $\Vert x\Vert _p$ represents the standard $\ell _p$
norm of $x$, and $\Vert x\Vert _0$ denotes the cardinality of $%
\mathop{\mathrm{supp}}(x)$, that is the number of nonzero entries of $x$.

For a matrix $X\in {{\mathbb R}^{N,M}}$ (with $M$ possibly equal to $\infty $%
), we denote by $[X]_{i,j}$ the entry of $X$ in row $i$ and column~$j.$ For $%
n\leq M$, we denote by $X_{\uparrow n}\in {{\mathbb R}^{N,n}}$ the
sub-matrix formed by the first $n$ columns of $X$, with $X_{\downarrow n}\in
{{\mathbb R}^{N,M-n}}$ the sub-matrix formed by the columns of $X$ of
indices $n+1,\ldots ,M$, and with $X_{\sharp n}$ the $n\times n$ principal
sub-matrix of $X$. The identity matrix is denoted by $I$, or by $I_n$, if we
wish to specify its dimension. We denote by $X^{\dagger }$ the Moore-Penrose
pseudo-inverse of $X$; if $X$ has full column rank, then $X^{\dagger
}=(X^{\top }X)^{-1}X^{\top }$.

If $x$ is a random variable, then $\mathbb{E}\{x\}$ denotes the expected
value of $x$, and ${\mathrm{var}} \{x\}$ denotes its variance: ${\mathrm{var}%
} \{x\} = \mathbb{E} \{ (x- \mathbb{E}\{x\})^2\}$. $\mathbb{P}$ denotes a
probability measure on $x$. The symbol $\leadsto $ implies almost sure
convergence, and it is formally defined in Section~\ref{sec:convergencesym}.

\subsection{Chebyshev's inequality for certain empirical means}

\label{sec:cheby} Let $x_i$, $i=1,\ldots ,$ be a sequence of (not
necessarily independent) random variables such that $\mathbb{E}\{x_i\}=\mu
<\infty $ for all $i$, ${\mathrm{var}}\{x_i\}=\sigma _i^2\leq \overline{%
\sigma }^2<\infty $ for all $i$, and $\mathbb{E}\{(x_i-\mu )(x_j-\mu )\}=0$
for all $i\neq j$. For given $N\geq 1$, define the empirical mean

\noindent
\[
\hat{x}_N\doteq \tfrac 1N\tsum\nolimits_{i=1}^Nx_i.
\]

\vspace*{-0.2em}

\noindent Obviously, from linearity of the expectation, it holds that $%
\mathbb{E}\{\hat{x}_N\}=\mu $. Further, we have that\vspace*{-1.2em}

\noindent
\begin{eqnarray*}
\lefteqn{\sigma ^2\!\doteq \mathrm{var}\{\hat{x}_N\}\!=\!\mathbb{E}\!\left\{
\!(\hat{x}_N\!-\!\mu )^2\right\} \!=\!\frac 1{N^2}\mathbb{E}\!\left\{
\!\left[ \tsum\nolimits_{i=1}^N(x_i\!-\!\mu )\!\right] ^{\!2}\right\} } \\
&=&\!\frac 1{N^2}\!\!\left[ \TeXButton{rule}{\rule[0mm]{0.0mm}{5mm}}%
\hspace*{-0.5mm}\right. \tsum_{i=1}^N\!\mathbb{E}\!\left\{ \!(x_i\!-\!\mu
)^2\right\} \!+\!\!\tsum_{i=1}^N\tsum_{j=1,\,j\neq i}^N\!\mathbb{E}\left\{
(x_i\!-\!\mu )(x_j\!-\!\mu )\right\} \left. \TeXButton{rule}
{\rule[0mm]{0.0mm}{5mm}}\hspace*{-1mm}\right] \\
&=&\tsum\nolimits_{i=1}^N\left. \sigma _i^2\right/ N^2\leq \left. \overline{%
\sigma }^2\right/ N,
\end{eqnarray*}

\vspace*{-1.1em}

\noindent where the last passages follow from the fact that the $x_i$s are
uncorrelated, and have first moment $\mu $ and variance $\sigma _i^2\!\leq \!%
\overline{\sigma }^2$. Chebyshev's inequality applied to the random
variable\thinspace $\hat{x}_N$\thinspace thus states that,\thinspace for any
$\eta \!>\!0$,

\noindent
\begin{equation}
\mathbb{P}\{|\hat{x}_N-\mu |\geq \eta \sigma \}\leq {1}/{\eta ^2}.
\label{eq:cheby}
\end{equation}

\vspace*{-0.35em}

\noindent Since $\eta \sigma \leq \eta \overline{\sigma }/\sqrt{N}$, we have
that $\mathbb{P}\{|\hat{x}_N-\mu |\geq \eta \overline{\sigma }/\sqrt{N}%
\}\leq \mathbb{P}\{|\hat{x}_N-\mu |\geq \eta \sigma \}$, whence, from (\ref
{eq:cheby}), we obtain that $\mathbb{P}\{|\hat{x}_N-\mu |\geq \eta \overline{%
\sigma }/\sqrt{N}\}\leq {1}/{\eta ^2}$. Equivalently, we can state that, for
any $\epsilon >0$, it holds that

\noindent
\[
\mathbb{P}\{|\hat{x}_N-\mu |\geq \epsilon \}\leq \left. \overline{\sigma }%
^2\right/ \left( N\epsilon ^2\right) .\label{eq:cheby:emp_mean}
\]

\vspace*{-0.35em}

\noindent We thus conclude that, for any given accuracy $\epsilon >0$ and
probability $\beta \in (0,1)$, it holds that

\noindent
\[
\mathbb{P}\{|\hat{x}_N-\mu |\geq \epsilon \}\leq \beta ,\quad \forall N\geq
\left\lceil \left. \overline{\sigma }^2\right/ \left( \beta \epsilon
^2\right) \right\rceil .\label{eq:cheby:emp_meanconv}
\]

\vspace*{-0.35em}

\noindent Notice that (\ref{eq:cheby}) implies that $\mathbb{P}\{|\hat{x}%
_N-\mu |>\eta \sigma \}\leq {1}/{\eta ^2}$; hence, by considering the
complementary event, it also holds that $\mathbb{P}\{|\hat{x}_N-\mu |\leq
\eta \sigma \}\geq 1-{1}/{\eta ^2}$, from which it follows that

\noindent
\[
\mathbb{P}\{|\hat{x}_N-\mu |\leq \epsilon \}\geq 1-\left. \overline{\sigma }%
^2\right/ \left( N\epsilon ^2\right) .\label{eq:cheby:emp_meanconv_comp}
\]

\subsubsection{Meaning of the convergence symbol $\leadsto $}

\label{sec:convergencesym} For a random variable $z_N$ that depends on $N$
and for a given real value $\bar{z}$, the notation $z_N\leadsto \bar{z}$
means that for any given $\epsilon >0$ and $\beta \in (0,1)$ there exists a
\emph{finite} integer $N_{\epsilon ,\beta }$ such that

\noindent
\begin{equation}
\mathbb{P}\{|z_N-\bar{z}|\geq \epsilon \}\leq \beta ,\quad \forall N\geq
N_{\epsilon ,\beta }.  \label{eq:convergence_def}
\end{equation}

\vspace*{-0.35em}

\noindent Notice that $z_N\leadsto \bar{z}$ implies that $z_N$ converges to $%
\bar{z}$ \emph{almost surely} (that is, with probability one), as $N$ tends
to infinity. However, we are specifically interested in the
property\thinspace in\hspace*{0.5mm}(\ref{eq:convergence_def}),%
\hspace*{0.5mm}that\thinspace holds\thinspace for\thinspace
possibly\thinspace large,\hspace*{0.5mm}but\thinspace finite,\hspace*{0.5mm}$%
N\!$.

\subsection{Lipschitz functions of random variables}

\label{sec:Lipschitz} If $z_N$ is the empirical mean of $N$ uncorrelated
variables with common mean $\mu $ and variance bounded by $\overline{\sigma }%
^2$ then, from the discussion in Section~\ref{sec:cheby}, we conclude that
indeed $z_N\leadsto \mu $ and, in particular, (\ref{eq:convergence_def})
holds for $N_{\epsilon ,\beta }=\left\lceil \left. \overline{\sigma }%
^2\right/ \left( \beta \epsilon ^2\right) \right\rceil $. However, we shall
use the convergence notation $z_N\leadsto \bar{z}$ also when $\bar{z}$ is
not necessarily the expected value of $z_N$, and/or when $z_N$ is not
necessarily an empirical mean. The following lemma holds.\vspace*{0.3em}

\begin{lemma}
\label{lem:lip_conv} For any fixed integer $p$, let $y_1,\ldots ,y_p$ be
(possibly correlated) scalar random variables that depend on $N$ and such
that $y_i\leadsto \bar{y}_i$, $i=1,\ldots ,p$, for some given values $\bar{y}%
_1,\ldots ,\bar{y}_p$. Let $f$ be a Lipschitz continuous function from ${\ {%
\mathbb R}^p}$ into ${\ {\mathbb R}^{}}$, such that $f(\bar{y}_1,\ldots ,%
\bar{y}_p)$ is finite. Then, it holds that $f(y_1,\ldots ,y_p)\leadsto f(%
\bar{y}_1,\ldots ,\bar{y}_p)$.\vspace*{-1.0em}

\TeXButton{qed}
{\hbox to \hsize{\hfill \vrule height 1.6ex width
1.5ex depth -.1ex}}\vspace*{0.3em}
\end{lemma}

\noindent
Appendix~\ref{sec:app_proof:lip_conv} contains a proof of Lemma \ref
{lem:lip_conv}.%

\section{Problem setup}

\label{sec:setup}

\subsection{A linear measurement model}

\label{sec:model} We consider an identification experiment in which a
discrete-time scalar input signal $\tilde{u}(k)$ enters an LTI dynamic
system, which produces in response a scalar output signal $\tilde{y}(k)$.
This output is acquired via noisy measurements over a time window $%
k=1,\ldots ,N$, obtaining a sequence of output measurements $y(k)=\tilde{y}%
(k)+\delta _y(k)$, $k=1,\ldots ,N$, where $\delta _y(k)$ is the measurement
noise sequence. Since the unknown system is assumed to be LTI, there exists
a linear relation between the output measurements and the unknown system's
impulse response $h(i)$, $i=1,\ldots $ Assuming that the system is operating
in steady state, this relation is given by the discrete-time convolution:
for $k=1,\ldots,N$,

\noindent
\begin{equation}
y(k)\!=\!\tilde{y}(k)+\delta _y(k)\!=\!\!\tsum_{i=1}^\infty \!\tilde{u}%
(k-i+1)h(i)+\delta _y(k).  \label{eq:yk_inf}
\end{equation}

\vspace*{-0.2em}

\noindent Observe that, following a nonparametric approach, we do not assume
to know in advance the order of the unknown system; therefore, in (\ref
{eq:yk_inf}), all values $h(i)$ can be, a priori, nonzero. Letting

\noindent
\[
y\!\doteq \!\left[ \!
\begin{array}{c}
\vspace*{-7.0mm} \\
y(1)\vspace*{-2.0mm} \\
y(2)\vspace*{-2.5mm} \\
\vdots \vspace*{-2.5mm} \\
y(N)\vspace*{-0.5mm}
\end{array}
\!\right] \!;\;\;\delta _y\!\doteq \!\left[ \!
\begin{array}{c}
\vspace*{-7.0mm} \\
\delta _y(1)\vspace*{-2.0mm} \\
\delta _y(2)\vspace*{-2.5mm} \\
\vdots \vspace*{-2.5mm} \\
\delta _y(N)\vspace*{-0.5mm}
\end{array}
\!\right] \!;\;\;\tilde{u}_i\!\doteq \!\left[ \!
\begin{array}{c}
\vspace*{-7.0mm} \\
\tilde{u}(2-i)\vspace*{-2.0mm} \\
\tilde{u}(3-i)\vspace*{-2.5mm} \\
\vdots \vspace*{-2.5mm} \\
\tilde{u}(N+1-i)\vspace*{-0.5mm}
\end{array}
\!\right] \!,
\]

\vspace*{-0.3em}

\noindent for $i=1,2,\ldots ,$ we can write (\ref{eq:yk_inf}) in vector
format as

\noindent
\begin{equation}
y=\tsum\nolimits_{i=1}^\infty \tilde{u}_ih(i)+\delta _y.
\label{eq:yk_inf_vec}
\end{equation}

\vspace*{-0.3em}

\noindent For any integer $n\geq 0$, we define

\noindent
\[
\tilde{U}_{\uparrow n}\doteq [\tilde{u}_1\,\cdots \,\tilde{u}_n]\in {\!{%
\mathbb R}^{N,n}},\;\;h_{\uparrow n}\doteq [h(1)\,\cdots \,h(n)]^{\top }{\!}%
\in {\!{\mathbb R}^n},
\]

\vspace*{-0.4em}

\noindent as well as the semi-infinite matrices and vectors

\noindent
\[
\begin{array}{l}
\tilde{U}_{\downarrow n}\doteq [\tilde{u}_{n+1}\;\tilde{u}_{n+2}\;\cdots
]\in {\!{\mathbb R}^{N,\infty }},\TeXButton{rule}{\rule[-2.5mm]{0mm}{2.5mm}}%
\vspace*{-0.5em} \\
h_{\downarrow n}\doteq [h(n+1)\;h(n+2)\;\cdots ]^{\top }{\!}\in {\!{\mathbb R%
}^\infty }.
\end{array}
\]

\vspace*{-0.4em}

\noindent Let now $q\leq N$ be a given integer: our goal is to estimate the
first $q$ elements of the impulse response $h$ (i.e., to estimate $%
h_{\uparrow q}\in {{\mathbb R}^q}$), from $N$ noisy output measurements. The
value of $q$ is fixed by the decision maker, based on the available number
of measurements $N$ and on a priori knowledge. For instance, under a
standard assumption of stability (see Assumption~\ref{ass:main_stable}),
since $h(i)$ decays exponentially, one may a priori assess that the response
will be negligible for $i\geq q$, for some sufficiently large $q$. We can
then rewrite (\ref{eq:yk_inf_vec}) as

\noindent
\[
y=\tilde{U}_{\uparrow q}h_{\uparrow q}+\delta _y+y{^{\mathrm{ud}}},
\]

\vspace*{-0.4em}

\noindent where

\noindent
\[
y{^{\mathrm{ud}}}\doteq \tilde{U}_{\downarrow q}h_{\downarrow q}
\]

\vspace*{-0.3em}

\noindent represents the unmodelled dynamics due to the truncation of the
impulse response to the $q$-th term. For simplifying the notation, we let
from now on

\noindent
\[
\tilde{U}\doteq \tilde{U}_{\uparrow q},
\]

\vspace*{-0.3em}

\noindent which is an $N\times q$ Toeplitz matrix. \vspace{-0em}

\subsection{An Elastic Net identification criterion}

\label{sec:elnet} The initial approach that we consider for identifying the
unknown system's impulse response consists in finding an estimate of $%
h_{\uparrow q}$ that minimizes w.r.t.\ $x$ the cost function\vspace*{-1.2em}

\noindent
\begin{eqnarray}
\hspace*{25mm}\frac 1\gamma \Vert y-\tilde{U}x\Vert _2^2+\Vert x\Vert _0,
\label{eq:cost1_prel}
\end{eqnarray}

\vspace*{-1.5em}

\noindent where $\gamma >0$ is a suitable tradeoff parameter. The first term
in (\ref{eq:cost1_prel}) is the standard prediction error, while the second
term $\Vert x\Vert _0$ represents the cardinality of $x$, that is the number
of nonzero entries in $x$. This term penalizes the complexity of the
estimate, thus promoting solutions with a small number of nonzero entries.
Note incidentally that, if $\delta _y(k)$ is a sequence of independent
identically distributed (i.i.d.)\ Normal random variables with zero mean and
known variance $\sigma _y^2$ then, for $\gamma =2\sigma _y^2$, the above
criterion coincides with the well-known Akaike's information criterion AIC.
Other standard criteria, such as the BIC, can also be obtained for different
values of $\gamma $.

\subsubsection{Input uncertainty and averaged cost}

\label{sec:input_cost} In a realistic identification experiment, however,
the input signal $\tilde{u}(k)$ that enters the unknown system is a possibly
``perturbed'' version of a nominal input signal $u(k)$ that the user intends
to provide to the system. To model this situation, we assume that $\tilde{u}%
(k)=u(k)+\delta _u(k)$, where $u(k)$ is the nominal input signal, and $%
\delta _u(k)$ is an i.i.d.\ random noise sequence, which is assumed to have
zero mean and variance $\sigma _u^2$ (setting $\sigma _u^2=0$ we recover the
standard, no-input-noise, situation). Considering the time window $%
k=1,\ldots ,N$, we have in matrix form that

\noindent
\begin{equation}
\tilde{U}=U+\Delta ,  \label{eq_input_noise}
\end{equation}

\vspace*{-0.4em}

\noindent where $U$ is an $N\times q$ Toeplitz matrix containing the nominal
input signal, and $\Delta $ is an $N\times q$ Toeplitz matrix containing the
noise samples $\delta _u(k)$. Specifically, $U\doteq [u_1\,\cdots \,u_q]$,
and $\Delta \doteq [\delta _1\,\cdots \,\delta _q]$, where for $i=1,\ldots
,q $

\noindent
\[
u_i\doteq \left[
\begin{array}{c}
\vspace*{-7.0mm} \\
u(2-i)\vspace*{-2.0mm} \\
u(3-i)\vspace*{-3.0mm} \\
\vdots \vspace*{-2.0mm} \\
u(N+1-i)\vspace*{-0.5mm}
\end{array}
\right] ,\quad \delta _i\doteq \left[
\begin{array}{c}
\vspace*{-7.0mm} \\
\delta _u(2-i)\vspace*{-2.0mm} \\
\delta _u(3-i)\vspace*{-3.0mm} \\
\vdots \vspace*{-2.0mm} \\
\delta _u(N+1-i)\vspace*{-0.5mm}
\end{array}
\right] .
\]

\vspace*{-0.3em}

\noindent We account for input uncertainty in the identification experiment
by ``averaging'' the effect of this uncertainty in the cost criterion (\ref
{eq:cost1_prel}). This leads to the following cost function:\vspace*{-1.2em}

\noindent
\begin{eqnarray}
\hspace*{5mm}J_0(x) &=&\mathbb{E}_{\delta _u}\left\{ \frac 1\gamma \Vert y-%
\tilde{U}x\Vert _2^2+\Vert x\Vert _0\right\}  \nonumber \\
&=&\frac 1\gamma \mathbb{E}_{\delta _u}\left\{ \Vert y-(U+\Delta )x\Vert
_2^2\right\} +\Vert x\Vert _0,  \label{eq:cost1}
\end{eqnarray}

\vspace*{-1.1em}

\noindent where $\mathbb{E}_{\delta _u}$ denotes expectation w.r.t.\ the
random sequence $\delta _u$. Elaborating on the expression (\ref{eq:cost1}),
we obtain\vspace*{-1.2em}

\noindent
\begin{eqnarray*}
\lefteqn{\mathbb{E}_{\delta _u}\{\Vert y-(U+\Delta )x\Vert _2^2\}} \\
\hspace*{5mm} &=&\mathbb{E}_{\delta _u}\{\Vert y-Ux\Vert _2^2+\Vert \Delta
x\Vert _2^2-2(y-Ux)^{\top }\Delta x\} \\
&=&\Vert y-Ux\Vert _2^2+\mathbb{E}_{\delta _u}\{\Vert \Delta x\Vert _2^2\} \\
&=&\Vert y-Ux\Vert _2^2+x^{\top }\mathbb{E}_{\delta _u}\{\Delta ^{\top
}\Delta \}x,
\end{eqnarray*}

\vspace*{-1.0em}

\noindent because $\mathbb{E}_{\delta _u}\{\Delta \}=0$. Since $\delta _u(k)$
is an i.i.d.\ sequence, and since $\Delta $ has Toeplitz structure, it is
easy to verify that the off-diagonal terms in $\mathbb{E}_{\delta
_u}\{\Delta ^{\top }\Delta \}$ are zero, while the diagonal terms are all
equal to $N\sigma _u^2$. Therefore, it holds that $\mathbb{E}_{\delta
_u}\{\Delta ^{\top }\Delta \}=N\sigma _u^2I_q$, and the expected cost $%
J_0(x) $ is explicitly expressed as

\noindent
\begin{equation}
J_0(x)=\frac 1\gamma \Vert y-Ux\Vert _2^2+\frac{N\sigma _u^2}\gamma \Vert
x\Vert _2^2+\Vert x\Vert _0.  \label{eq:exp_Akaike}
\end{equation}

Notice that this setting can be easily extended to wide-sense stationary
input noise sequences $\delta _u(k)$, in which case the second term in the
above expression takes the form $\frac N\gamma x^{\top }R_ux$, where $R_u$
is the autocorrelation matrix of $\delta _u$. For simplicity, however, we
here focus on the basic case of an i.i.d.\ sequence, for which $R_u=\sigma
_u^2I_q$. Observe further that accounting for noise on the input signal
results in the introduction of a Tikhonov-type regularization term in (\ref
{eq:exp_Akaike}), a fact that has been previously observed in other contexts
such as neural network training, see, e.g., \cite{Bishop:95}.

\subsubsection{Normalizing the variables}

\label{sec:normvars} We next rescale the variables in the cost (\ref
{eq:exp_Akaike}) by normalizing the columns of the regression matrix. First,
we rewrite $J_0(x)$ as\vspace*{-1.2em}

\noindent
\begin{eqnarray}
\hspace*{20mm}J_0(x)=\frac 1\gamma \left\| b-\bar{A}x\right\| _2^2+\Vert
x\Vert _0,  \label{eq:cost_manip_1}
\end{eqnarray}

\vspace*{-1.0em}

\noindent where

\noindent
\begin{equation}
b\doteq \left[
\begin{array}{c}
\vspace*{-7.0mm} \\
y\vspace*{-1.5mm} \\
0\vspace*{-0.5mm}
\end{array}
\right] ,\quad \bar{A}\doteq \left[
\begin{array}{c}
\vspace*{-7.0mm} \\
U\vspace*{-1.5mm} \\
\sigma _u\sqrt{N}I_q\vspace*{-0.5mm}
\end{array}
\right] .  \label{eq:bdef}
\end{equation}

\vspace*{-0.3em}

\noindent Second, we let $T\doteq \mbox {\rm diag}(\Vert \bar{a}_1\Vert
_2,\ldots ,\Vert \bar{a}_q\Vert _2)^{-1}$, where $\bar{a}_i$ denotes the $i$%
-th column of $\bar{A}$, and perform the change of variable $\tilde{x}%
=T^{-1}x$, thus the right-hand side of (\ref{eq:cost_manip_1}) becomes

\noindent
\begin{equation}
\tilde{J}_0(\tilde{x})\doteq \frac 1\gamma \Vert b-A\tilde{x}\Vert
_2^2+\Vert \tilde{x}\Vert _0,  \label{eq:normalized_cost}
\end{equation}

\vspace*{-0.3em}

\noindent where we defined $A\doteq \bar{A}T$, and we used the fact that $%
\Vert T\tilde{x}\Vert _0=\Vert \tilde{x}\Vert _0$, since the cardinality of
a vector does not depend on (nonzero) scalings of the entries of the vector.
We observe that the columns $a_1,\ldots ,a_q$ of $A$ now have unit Euclidean
norm. We let $\tilde{x}_0^{*}\doteq \arg \min \,\tilde{J}_0(\tilde{x})$, and
$x_0^{*}\doteq \arg \min \,J_0(x)$, where it obviously holds that $x_0^{*}=T%
\tilde{x}_0^{*}$. These optimal solutions are hard to determine numerically
in practice. However, we do not need to compute them, we only need them for
theoretical purposes.

\subsubsection{Weighted $\ell _1$ relaxation of the cost function}

\label{sec:relaxation} We now introduce the following tractable relaxation
of the cost (\ref{eq:normalized_cost}):

\noindent
\begin{equation}
\tilde{J}_1(\tilde{x})\doteq \frac 1\gamma \Vert b-A\tilde{x}\Vert
_2^2+\Vert W\tilde{x}\Vert _1.  \label{eq:normalized_cost_l1}
\end{equation}

\vspace*{-0.3em}

\noindent {where }$W\doteq \limfunc{diag}(w_1,\ldots ,w_q)$ is a suitable
weighting matrix, with $\max_{k=1,\ldots ,q}w_k=1$, $\min_{k=1,\ldots
,q}w_k>0$. We shall henceforth assume that the weight sequence is
nondecreasing: $w_1\leq w_2\leq \cdots \leq w_q=1$.

Notice that, expanding the squared norm in (\ref{eq:normalized_cost_l1}), we
obtain the cost function $\tilde{J}_1$ in the form\vspace*{-0.1em}

\noindent
\begin{equation}
\tilde{J}_1(\tilde{x})=\frac 1\gamma \Vert y-UT\tilde{x}\Vert _2^2+\frac{%
N\sigma _u^2}\gamma \Vert T\tilde{x}\Vert _2^2+\Vert W\tilde{x}\Vert _1,
\label{eq:gen_elnettil}
\end{equation}

\vspace*{-0.3em}

\noindent which corresponds to the cost expressed in the original variable $%
x=T\tilde{x}$\vspace*{-0.1em}

\noindent
\begin{equation}
J_1(x)=\frac 1\gamma \Vert y-Ux\Vert _2^2+\frac{N\sigma _u^2}\gamma \Vert
x\Vert _2^2+\Vert WT^{-1}x\Vert _1.  \label{eq:gen_elnet}
\end{equation}

\vspace*{-0.2em}

\noindent The cost function (\ref{eq:gen_elnettil}) is strongly convex,
hence the optimal solution $\tilde{x}_1^{*}\doteq \arg \min \,\tilde{J}_1(%
\tilde{x})$ is unique and, equivalently, the minimization of (\ref
{eq:gen_elnet}) has a unique optimal solution $x_1^{*}=T\tilde{x}_1^{*}$. In
the following section, we shall study the properties of $x_1^{*}$ as an
estimate of the impulse response $h_{\uparrow q}$. Note that only two
parameters ($\gamma $ and $\sigma _u$) have to be chosen to obtain this
estimate. A systematic procedure is proposed in Section \ref{sec:id_proc},
allowing an effective choice of these parameters, based on the desired
trade-off between model complexity and accuracy.

\begin{remark}
\TeXButton{rm}{\rm}The cost criterion appearing in (\ref{eq:gen_elnettil})
is a particular version of the Lasso (see, e.g., \cite{Tibshirani:96}),
known as the Elastic Net (\cite{ZouHas:05}). The Elastic Net criterion
includes an $\ell _2$ regularization term which provides shrinkage and
improves conditioning of the $\ell _2$-error cost (by guaranteeing strong
convexity of the cost), as well as an $\ell _1$ penalty term which promotes
sparsity in the solution. Elastic Net-based methods are widely used in
statistics and machine learning, see, e.g., \cite{DeDeRo:09,HaTiFr:09}, and
are amenable to very efficient large-scale solution algorithms (\cite
{FrHaTi:10}). To the best of the authors' knowledge, this is the first work
in which the Elastic Net criterion is used in the context of a system
identification problem and the resulting sparsity pattern is rigorously
analyzed.
\end{remark}

\section{Leading response recovery}

\label{sec:lrr} This section contains the main results of the paper. First,
we report a preliminary technical lemma (Lemma~\ref{lem:recovery_1}) stating
that, under a certain condition, the minimizer $x_1^{*}$ of (\ref
{eq:gen_elnet}) is supported on $\{1,\ldots ,n\}$, with $n\leq q$. Second,
under some suitable assumptions on the input and noise signals, we show
(Theorem~\ref{thm:main}) that if the unknown system is stable, then for a
sufficiently large $N$ and for a given $n\leq q$, there exist explicitly
given $\gamma $ values for which the support of $x_1^{*}$ is contained in $%
\{1,\ldots ,n\}$, with any given high probability. This means that the
estimated impulse response $x_1^{*}$ is not only sparse but, with high
probability, it is zero precisely on the tail of the system's impulse
response $h_{\uparrow q}$. We next define the notions of \emph{leading
response} and \emph{leading support} of the system's impulse response, and
show (Corollary~\ref{cor:main_lead}) that if the unknown system is stable,
then for a suitable $\gamma $ and a sufficiently large $N$ the support of $%
x_1^{*}$ is contained in the leading support, with any given high
probability; we call this property \emph{leading response recovery} (%
\mbox{LRR}). Finally, we show (Corollary~\ref{cor:main_fir}) that if the
true unknown system is FIR then, for a sufficiently large $N$ and for any $%
\gamma >0$, the estimated impulse response $x_1^{*}$ will be sparse, and of
order no larger than the order of the true system, with high probability.

\subsection{Preliminary results, assumptions and definitions}

\label{sec:lrr_preliminary}With the notation set in Section~\ref
{sec:normvars}, for a given integer $n\leq q$, let $P_n\doteq A_{\uparrow
n}A_{\uparrow n}^{\dagger }$ denote the orthogonal projector onto the span
of $A_{\uparrow n}$, and define the \emph{$n$-leading recovery coefficient} $%
\Upsilon _n(A)\doteq 1-\max_{n<i\leq q}w_i^{-1}\left\| W_{\sharp
n}A_{\uparrow n}^{\dagger }a_i\right\| _1$, where $a_i$ is the $i$-th column
of $A$. The following technical lemma, based on a result in \cite{Tropp:06},
holds.\vspace*{0.3em}

\begin{lemma}
\label{lem:recovery_1} Suppose that for some integer $n\leq q$ it holds that

\noindent
\begin{equation}
\left\| W^{-1}A^{\top }(b-P_nb)\right\| _\infty \leq \gamma \left. \Upsilon
_n(A)\right/ 2,  \label{eq:lem_condition}
\end{equation}

\vspace*{-0.2em}

\noindent and let $x_1^{*}$ be the minimizer of (\ref{eq:gen_elnet}). Then,
it holds that

\noindent
\[
\mathop{\mathrm{supp}}(x_1^{*})\subseteq \{1,\ldots ,n\}.
\]

\vspace*{-1.2em}\TeXButton{qed}
{\hbox to \hsize{\hfill \vrule height 1.6ex width
1.5ex depth -.1ex}}\vspace*{0.0em}
\end{lemma}

\noindent
See Appendix~\ref{sec:app_proof:lem1} for a proof of Lemma~\ref
{lem:recovery_1}.

Let us now state the following working assumptions.\vspace*{0.2em}

\begin{assumption}[\it{on input and disturbance sequences\thinspace}\bf{}]
\hspace*{0mm} \\[-1.0em]\label{ass:main}

\begin{enumerate}
\item  The input $u(k)$ is an i.i.d.\ sequence with zero mean, bounded
variance $\nu ^2$ and bounded $4$-th order moment $\mathbb{E}\left\{
u(k)^4\right\} =\overline{m\TeXButton{rule}{\rule{0mm}{2mm}}_4}\,$.

\item  The noise $\delta _y(k)$ is an i.i.d.\ sequence with zero mean and
bounded variance $\sigma _y^2\,$.

\item  The input perturbation $\delta _u(k)$ is an i.i.d.\ sequence with
zero mean and bounded variance $\sigma _u^2\,$.

\item  $u(k)$, $\delta _y(k)$, and $\delta _u(k)$ are mutually uncorrelated. %
\vspace*{0.3em}
\end{enumerate}
\end{assumption}

\begin{assumption}[\it{Stability\thinspace}\bf{}]
\label{ass:main_stable} The unknown system's impulse response $h$ is such
that $|h(i)|\leq L\rho ^{i-1}$, for $i=1,2,\ldots ,$ for some given finite $%
L>0$ and $\rho \in (0,1)$.\vspace*{0.3em}
\end{assumption}

\noindent
We next establish a preliminary lemma.\vspace*{0.3em}

\begin{lemma}
\label{lem:convergence}Under Assumption~\ref{ass:main}, for any pair of
column vectors $u_i$ and $u_j$ it holds that\vspace*{-0.1em}

\noindent
\begin{equation}
\tfrac 1Nu_i^{\top }u_j\leadsto \left\{
\begin{array}{l}
\vspace*{-7.5mm} \\
\nu ^2\quad \mbox{if }i=j\vspace*{-2.0mm} \\
0\quad \mbox{otherwise,}\vspace*{-1.5mm}
\end{array}
\right.   \label{lem:convergence_claima}
\end{equation}

\vspace*{-0.2em}

\noindent where the notation $\leadsto $ has the meaning specified in
Section~\ref{sec:convergencesym}. Also, it holds that\vspace*{-1.0em}

\noindent
\begin{eqnarray}
\hspace*{22mm}\tfrac 1Nu_i^{\top }\delta _y\leadsto 0,\quad \forall i
\label{lem:convergence_claimb} \\
\hspace*{25mm}\tfrac 1Nu_i^{\top }\delta _j\leadsto 0,\quad \forall i,j.
\label{lem:convergence_claimc}
\end{eqnarray}

\vspace*{-1.1em}\TeXButton{qed}
{\hbox to \hsize{\hfill \vrule height 1.6ex width
1.5ex depth -.1ex}}\vspace*{-0.5em}
\end{lemma}

\noindent
Appendix~\ref{sec:app_proof:lem_convergence} contains a proof of Lemma~\ref
{lem:convergence}.%

We next define the notion of \emph{leading order} of the system's impulse
response, and the associated notions of \emph{leading response} and \emph{%
leading support}.\vspace*{0.3em}

\begin{definition}
\label{def:leading_response} Let Assumption~\ref{ass:main_stable} hold. We
define the \emph{leading order}, $n_l(N)$, of $h$ as the largest integer $%
i\leq q$ such that

\noindent
\begin{equation}
L\rho ^{i-1}\geq \frac{\sigma _y}\nu \times \frac 1{\sqrt{N}}.
\label{eq:leadingorder}
\end{equation}

\vspace*{-0.5em}

\noindent
The \emph{leading response} is $\{h(i),\,i=1,\ldots ,n_l\}$ and the \emph{%
leading support} is $\{1,\ldots ,n_l\}$.\vspace*{0.3em}
\end{definition}

\begin{remark}
\TeXButton{rm}{\rm}\label{rem:leading_order}We provide an intuitive
interpretation of the definition in (\ref{eq:leadingorder}). The leading
order is a value such that for time values larger than it the system's
impulse response cannot essentially be discriminated from noise. Indeed, if
a classical output error criterion would be used for estimating $h_{\uparrow
q}$, then the covariance matrix of the estimated parameter would be of the
form ${\sigma _y^2}(U^{\top }U)^{-1}$, which tends to ${\sigma _y^2}/({\nu
^2N})I_q$ as $N\to \infty $, see the proof of Theorem~\ref{thm:main} for
details. The standard error on the generic element $h(i)$ of the impulse
response thus goes to zero as\thinspace $1\!/\!\sqrt{\!N}$, where\thinspace $%
N$\thinspace is\thinspace the\thinspace number\thinspace of\thinspace
measurements\thinspace and\thinspace the proportionality\thinspace
constant\thinspace $\sigma _{\!y}/_{\!}\nu $\thinspace is\thinspace
the\thinspace noise-to-signal\thinspace ratio. \\ The leading order $n_l$ is
therefore defined as the time value after which the upper bound on $|h(i)|$
goes below the level $\eta =\frac{\sigma _y}\nu \frac 1{\sqrt{N}}$, and
hence $h(i)$ becomes essentially indistinguishable from noise, for all $i>n_l
$; if this condition is not met for $i\leq q$, then we just set $n_l=q$. It
is an immediate consequence of (\ref{eq:leadingorder}) that the leading
order grows as the logarithm of $N$, until it saturates to $q$:%
\vspace*{-0.2em}

\noindent
\[
n_l(N)=\min \left( \left\lfloor \frac{\log (\nu L)+\frac 12\log N-\log
(\sigma _{\!y}\rho )}{\log (\rho ^{-1})}\right\rfloor ,q\right) .\text{%
\vspace*{-0.2em}}
\]

\vspace{-0em}
\end{remark}

\subsection{Main results}

\label{sec:lrr_results} We next establish the main results of this paper.%
\vspace*{0.3em}

\begin{theorem}
\label{thm:main} Let Assumptions~\ref{ass:main} and \ref{ass:main_stable}
hold. Let $n\leq q$, $\kappa \doteq \nu /\sqrt{\nu ^2+\sigma _u^2}$, and%
\vspace*{-0.0em}

\noindent
\begin{equation}
\gamma =2\mu w_n^{-1}L\rho ^n\nu \kappa \times \sqrt{N}\text{,}
\label{eq:gamma_lower}
\end{equation}

\vspace*{-0.3em}

\noindent
for some $\mu >1$. Then, for any given $\beta \in (0,1)$ there exists a
finite integer $N_\beta $ such that for any $N\geq N_\beta $ it holds that%
\vspace*{-0.0em}

\noindent
\[
\mathop{\mathrm{supp}}(x_1^{*})\subseteq \{1,\ldots ,n\}
\]

\vspace*{-0.3em}

\noindent
with probability no smaller than $1-\beta $, where $x_1^{*}$ is the
minimizer of (\ref{eq:gen_elnet}).\vspace*{-1.0em}

\TeXButton{qed}
{\hbox to \hsize{\hfill \vrule height 1.6ex width
1.5ex depth -.1ex}}\vspace*{0.2em}
\end{theorem}

\noindent
Appendix~\ref{sec:app_proof:thm_main} contains a proof of Theorem~\ref
{thm:main}.
The key point of this theorem is that if the tradeoff parameter $\gamma $ is
chosen proportional to $\sqrt{N}$ then, with high probability and for a
sufficiently large $N$, the minimization of (\ref{eq:gen_elnet}) provides a
solution which is not only sparse, but its sparsity pattern is identically
zero on the tail of the impulse response, i.e., the estimated impulse
response $x_1^{*}$ is FIR of order at most $n$.

A consequence of Theorem~\ref{thm:main} is stated in the following
corollary: for a suitable constant value of $\gamma $, the minimizer $%
x_1^{*} $ of (\ref{eq:gen_elnet}) has its support contained in the leading
support.\vspace*{0.3em}

\begin{corollary}[\it{Leading support recovery\thinspace}\bf{}]
\label{cor:main_lead} \hspace*{0mm} \newline Let Assumptions~\ref{ass:main}
and~\ref{ass:main_stable} hold. Let $\kappa \doteq \nu /\sqrt{\nu ^2+\sigma
_u^2}$, and

\noindent
\begin{equation}
\gamma >2w_{n_l(N)}^{-1}\rho \sigma _y\kappa .  \label{eq:gamma_lower_const}
\end{equation}

\vspace*{-0.3em}

\noindent
Then, for any given $\beta \in (0,1)$ there exists a finite integer $N_\beta
$ such that for any $N\geq N_\beta $ it holds that

\noindent
\[
\mathop{\mathrm{supp}}(x_1^{*})\subseteq \{1,\ldots ,n_l(N)\}
\]

\vspace*{-0.3em}

\noindent
with probability no smaller than $1-\beta $, where $x_1^{*}$ is the
minimizer of (\ref{eq:gen_elnet}), and $n_l(N)$ is the leading order of the
unknown system's impulse response.\vspace*{-1.0em}

\TeXButton{qed}
{\hbox to \hsize{\hfill \vrule height 1.6ex width
1.5ex depth -.1ex}}\vspace*{0.3em}
\end{corollary}

\noindent
See Appendix~\ref{sec:app_proof:cor_main_lead} for a proof of Corollary~\ref
{cor:main_lead}.
Corollary~\ref{cor:main_lead} states that, under suitable conditions, an
estimate of the impulse response based on the minimization of (\ref
{eq:gen_elnet}) is supported inside the leading support of the system, with
high probability. The following corollary provides a similar result, for the
case in which the true system is a-priori known to have finite impulse
response (FIR).\vspace*{0.3em}

\begin{corollary}[\it{FIR recovery\thinspace}\bf{}]
\label{cor:main_fir} Let Assumption~\ref{ass:main} hold. Further, assume the
``true,'' unknown, system is FIR of order $n\leq q$, with $n$ unknown. Then,
for any $\gamma >0$ and for any given $\beta \in (0,1)$ there exists a
finite integer $N_\beta $ such that for any $N\geq N_\beta $ it holds that $%
\mathop{\mathrm{supp}}(x_1^{*})\subseteq \{1,\ldots ,n\}$ with probability
no smaller than $1-\beta $, where $x_1^{*}$ is the minimizer of (\ref
{eq:gen_elnet}).\vspace*{-1.0em}

\TeXButton{qed}
{\hbox to \hsize{\hfill \vrule height 1.6ex width
1.5ex depth -.1ex}}\vspace*{0.3em}
\end{corollary}

\noindent
See Appendix~\ref{sec:app_proof:cor_main} for a proof of Corollary~\ref
{cor:main_fir}.
The key point of this corollary is that if the true system is known to be
FIR, then the minimizer of (\ref{eq:gen_elnet}) will tendentially recover
the true order of the system, regardless of the value of $\gamma \!>\!0$
(but,\thinspace of course,\thinspace the larger the value of $\gamma $, the
sooner w.r.t.\thinspace $N$ the condition (\ref{eq:lem_condition}) will be
satisfied).

\section{Identification procedure}

\label{sec:id_proc}

We next formalize a possible procedure illustrating how the proposed
methodology can be used in a practical experimental setting. Suppose that a
set of data $\{y(k),u(k)\}_{k=3-q}^{N}$ is available from a process of the
form \eqref{eq:yk_inf}. Identification of the impulse response $h(i)$ is
performed by minimizing the cost function \eqref{eq:gen_elnet}. This
operation requires the choice of two parameters ($\gamma$ and $\sigma_u$).
If $\sigma_u$ and $\sigma_y$ are known from some a-priory information on the
noises affecting the system or can be reliably estimated, then $\gamma$ can
be chosen according to \eqref{eq:gamma_lower_const}, where $\rho$ can be
estimated by means of the technique in \cite{MiRT10} (see Section \ref
{sec:example1}). If instead this information is not available, a systematic
procedure for the choice of $\gamma$ and $\sigma_u$ is the following one:
\smallskip

\begin{itemize}
\item  Take ``reasonable'' sets $\Gamma =\{\gamma ^{(1)},\gamma ^{(2)},...\}$
and $\Sigma _u=\{\sigma _u^{(1)},\sigma _u^{(2)},...\}$ for $\gamma $ and $%
\sigma _u$ values, respectively. If $\sigma _u$ is known from some a-priory
information on the noise affecting the input, then $\Sigma _u=\sigma _u$.%
\vspace*{0.2em}

\item  Define $y$, $U$ and $T$ as shown in Section \ref{sec:setup}.%
\vspace*{0.2em}

\item  Run the following algorithm:
\[
\begin{array}{l}
\vspace*{-3.0em} \\
\text{for }i=1:\text{length}(\Sigma _u)\vspace*{-1.0mm} \\
\qquad \text{for }j=1:\text{length}(\Gamma )\vspace*{-1.0mm} \\
\qquad \qquad \sigma _u=\sigma _u^{(i)};\ \gamma =\gamma ^{(j)};%
\vspace*{-1.0mm} \\
\qquad \qquad x^{*}(i,j)=\arg \min_xJ_1(x);\vspace*{-1.0mm} \\
\qquad \qquad E(i,j)=\left\| y-Ux^{*}(i,j)\right\| _2^2;\vspace*{-1.0mm} \\
\qquad \qquad C(i,j)=\left\| x^{*}(i,j)\right\| _0;\vspace*{-1.0mm} \\
\qquad \text{end}\vspace*{-1.0mm} \\
\qquad \text{plot}(C(i,:),E(i,:))\text{{}}\vspace*{-1.0mm} \\
\text{end}\vspace*{-1.0mm}
\end{array}
\]
\vspace*{-1.0em}

\item  The obtained plot shows how the model accuracy (measured by $E$)
changes in function of its complexity (measured by $C$). Thus, $\gamma $ and
$\sigma _u$ can be chosen according to the desired trade-off between model
accuracy and complexity.\vspace*{0.2em}
\end{itemize}

Choosing $\gamma ^{(1)}>\gamma ^{(2)}>...$ and using $x^{*}(i,j-1)$ at the $j
$th step as the initial condition for the optimization problem may
significantly increase the speed of the algorithm. An example of application
of this procedure is shown in Section \ref{sec:example2} and, in particular,
in Figure~\ref{fig:Pareto}.

The weighting matrix $W$ plays a relevant role in the model order selection,
increasing the algorithm efficiency especially in situations where a low
number of data is available. For simplicity, unitary weights $w_i$ were
here adopted in Section \ref{sec:example}. Further research activity
will be devoted to investigate how to automatically and optimally select
these weights, in order to take into account possible priors on the
unknown system.

\section{Numerical examples}

\label{sec:example}

\subsection{A simulated LTI system}

\label{sec:example1} For our first numerical test we considered a classical
discrete-time LTI system proposed in \cite{Ljung:99_proc}. This system is
defined by the discrete-time transfer function

\noindent
\begin{equation}
H(z)=\frac{z^3+0.5z^2}{z^4-2.2z^3+2.42z^2-1.87z+0.7225},  \label{eq:true1}
\end{equation}

\vspace*{-0.2em}

\noindent
with sampling time $1$ s. We assume that all necessary parameters (e.g., the
noise variances and the impulse response's stability degree bounds) are
known or have been estimated in advance by other means.

\subsubsection{Experiments with a fixed number of data}

\label{sec:example1_fixed_N} Three i.i.d.\ input sequences with zero mean
and variance $\nu ^2=1$ were first generated. Each of these sequences was
corrupted by an i.i.d.\ noise with zero mean and variance $\sigma _u^2$,
with $\sigma _u=0.01$ for the first sequence, $\sigma _u=0.03$ for the
second one, and $\sigma _u=0.05$ for the third one. These values correspond
to noise-to-signal standard deviation ratios of $1\%$, $3\%$, and $5\%$,
respectively.

For each noise-corrupted input sequence, the system (\ref{eq:true1}) was
simulated for $2000$ s, assuming zero initial conditions. Note that the
system reaches steady-state conditions after about $150$ s. The resulting
output sequence was corrupted by an i.i.d.\ noise with zero mean and
variance $\sigma_{y}^{2}$, with $\sigma_{y}=0.1$ for the first sequence, $%
\sigma_{y}=0.3$ for the second one, and $\sigma_{y}=0.5$ for the third one.
These values correspond to noise-to-signal standard deviation ratios of $1\%$%
, $3\%$, and $5\%$, respectively (the system static gain is about $10$).
Then, the last $N=1000$ noise-corrupted output values were acquired. From
these data, the following models of the unknown system impulse response were
identified:

\begin{itemize}
\item  Leading Response Recovery (\mbox{LRR}) model. This model was obtained
minimizing the objective function (\ref{eq:gen_elnet}). The parameters
required for this minimization were taken as follows. The variances $\sigma
_u^2$ and $\sigma _y^2$ were assumed known (or accurately estimated). The
impulse response bound parameters were estimated by means of the technique
in \cite{MiRT10}, giving values $L=6$ and $\rho =0.93$ (note that only $\rho
$ is required by the \mbox{LRR} algorithm). {Unitary weights $w_i$ were
adopted. The estimated length was taken as $q=500$.} The value of $\gamma $
was chosen according to (\ref{eq:gamma_lower_const}).

\item  Least Squares (LS) model. This model was identified using standard
least squares, that is, by minimizing the objective function (\ref
{eq:gen_elnet}), with $\sigma _u^2=0$ and $T^{-1}=0$.

\item  Tikhonov regularized Least Squares (TLS) model. This model was
identified by minimizing the objective function (\ref{eq:gen_elnet}), with $%
T^{-1}=0$.
\end{itemize}

To validate the identified models, the following indices were computed:

\begin{itemize}
\item  Best fit criterion:

\noindent
\[
\mbox{FIT}\doteq 100\left( 1-\frac{\left\| y-\hat{y}\right\| _2}{\left\| y-%
\mathrm{mean}(y)\right\| _2}\right)
\]

\vspace*{-0.3em}

\noindent
where $y$ is the measured system output vector and $\hat{y}$ is the output
vector simulated by the model. The $\mbox{FIT}$ index was evaluated on $%
N_v=2000$ validation data points (i.e., points not previously used for
identification). Obviously, this index measures the model simulation
accuracy: the closer it is to $100\%$, the more accurate the simulation is.

\item  Tail $\ell _0$ quasi-norm:

\noindent
\[
\mbox{\mbox{\mbox{TN0}}}\doteq \left\| x_{\mathrm{tail}}^{*}\right\| _0
\]

\vspace*{-0.3em}

\noindent
where $x_{\mathrm{tail}}^{*}\doteq [x^{*}(n_l+1)\,\cdots \,x^{*}(500)]^{\top
}$ and $x^{*}$ is the estimated model impulse response. This index is a
measure of the model tail (the tail can be defined as the vector formed by
the impulse response components with index $>n_l$). More precisely, it
counts how many elements in the tail of the model impulse response are
different from zero. Note that, for $\sigma _y=0.1$, $n_l=105$; for $\sigma
_y=0.3$, $n_l=89$; for $\sigma _y=0.5$, $n_l=82$.

\item  Tail $\ell _1$ norm:

\noindent
\[
\mbox{TN1}\doteq \left\| x_{\mathrm{tail}}^{*}\right\| _1.
\]

\vspace*{-0.3em}

\noindent
This index provides an indication on the average magnitude of the elements
in the tail of the model impulse response.
\end{itemize}

A Monte Carlo simulation was then carried out, where the above
identification-validation procedure was repeated for $100$ trials. The
averages $\overline{\mbox{FIT}}$, $\overline{\mbox{\mbox{\mbox{TN0}}}}$ and $%
\overline{\mbox{TN1}}$ of $\mbox{FIT}$, $\mbox{\mbox{\mbox{TN0}}}$ and $%
\mbox{TN1}$ obtained in this simulation are reported in Table \ref{tab1}. We
observe that the three identification methods lead to very similar $%
\mbox{FIT}$ values. However, the $\mbox{LRR}$ models have a tail that is
practically null (in average, about $4$ non-null elements over about $280$),
even though the number of data used for identification is relatively low
(1000 data). This fact shows that our identification algorithm is able to
provide highly sparse models, without compromising their simulation
accuracy. An even more important aspect is that sparsification does not
occur for ``random'' indexes of the model impulse response but for large
indexes, i.e., those indexes associated with the exponentially decaying tail
of the impulse response.

\begin{table}[htbp]
\center
\begin{tabular}{|c|c||c|c|c|}
\hline
noise & model & $\overline{\mbox{FIT}}$ & $\overline{\mbox{\mbox{\mbox{TN0}}}%
}$ & $\overline{\mbox{TN1}}$ \\ \hline\hline
\multirow{3}{*}{1\%} & \mbox{LRR} & 98.6 & 6.0 & 0.012 \\ \cline{2-5}
& LS & 98.6 & 315 & 1.40 \\ \cline{2-5}
& TLS & 98.6 & 315 & 1.39 \\ \hline\hline
\multirow{3}{*}{3\%} & \mbox{LRR} & 95.9 & 4 & 0.019 \\ \cline{2-5}
& LS & 96.0 & 267 & 3.29 \\ \cline{2-5}
& TLS & 96.0 & 267 & 3.28 \\ \hline\hline
\multirow{3}{*}{5\%} & \mbox{LRR} & 93.3 & 3.3 & 0.025 \\ \cline{2-5}
& LS & 93.4 & 246 & 4.97 \\ \cline{2-5}
& TLS & 93.4 & 246 & 4.94 \\ \hline
\end{tabular}
\caption{Average indices obtained in the Monte Carlo simulation.}
\label{tab1}
\end{table}

It is important to remark that the \mbox{LRR} algorithm does not use the
prior information in terms of $L$ and $\rho $ values to impose strict
constraints or weights on the samples of the leading response. The
information on $L$ and $\rho $ is only used in the proof of Theorem~\ref
{thm:main} (see \eqref{lrho_ub}) to derive a bound on the value of $\gamma $
(see \eqref{eq:gamma_lower_const}).

It may be expected that using explicit constraints or weights based on $L$
and $\rho$ in the algorithm may lead to improvements in terms of model
accuracy and/or complexity. To better investigate this aspect, we performed
another Monte Carlo simulation, considering a 3\% noise level, and applying
standard constrained least squares and regularized Diagonal/Correlated
kernel methods (the latter using the Matlab routine \texttt{impulseest.m},
see, e.g., \cite{PDCD14}). Indeed, these methods use the $L$ and $\rho$
information (either known a priori or estimated from the data) to impose a
desired exponential decay of the overall impulse response. The following
index values were obtained with constrained least squares (CLS): $\overline{%
\mbox{FIT}}=96.6$, $\overline{\mbox{\mbox{\mbox{TN0}}}}=267$, $\overline{%
\mbox{TN1}}=0.115$. The following index values were obtained with the
regularized Diagonal/Correlated kernel method (DCK): $\overline{\mbox{FIT}}%
=96.7$, $\overline{\mbox{\mbox{\mbox{TN0}}}}=267$, $\overline{\mbox{TN1}}%
=0.033$.

We can compare these results with those shown in Table \ref{tab1}. It can be
noted that the CLS and DCK methods give slight improvements w.r.t.\ the
other methods in terms of the $\mbox{FIT}$ criterion, although the formers
use a significantly stronger prior information. An interesting result of the
CLS and DCK methods is that they lead to tails with very small (albeit
nonzero) elements, giving a relevant reductions of the tail magnitude
w.r.t.\ the LS and TLS methods, with $\overline{\mbox{TN1}}$ indexes not far
from the one given by the \mbox{LRR} method. Nevertheless, the $\overline{%
\mbox{\mbox{\mbox{TN0}}}}$ values given by the \mbox{LRR} method are by far
the lowest ones, showing that this method is the only one (among those
considered) allowing effective and unsupervised model order selection.

\subsubsection{Experiments with an increasing number of data}

\label{sec:example1_increasing_N} 
\begin{figure}[htb]
\vspace*{-0.2cm} \centering
\includegraphics[scale=0.45,bbllx=5mm,bblly=13mm,bburx=180mm,bbury=205mm,width=84mm,clip]{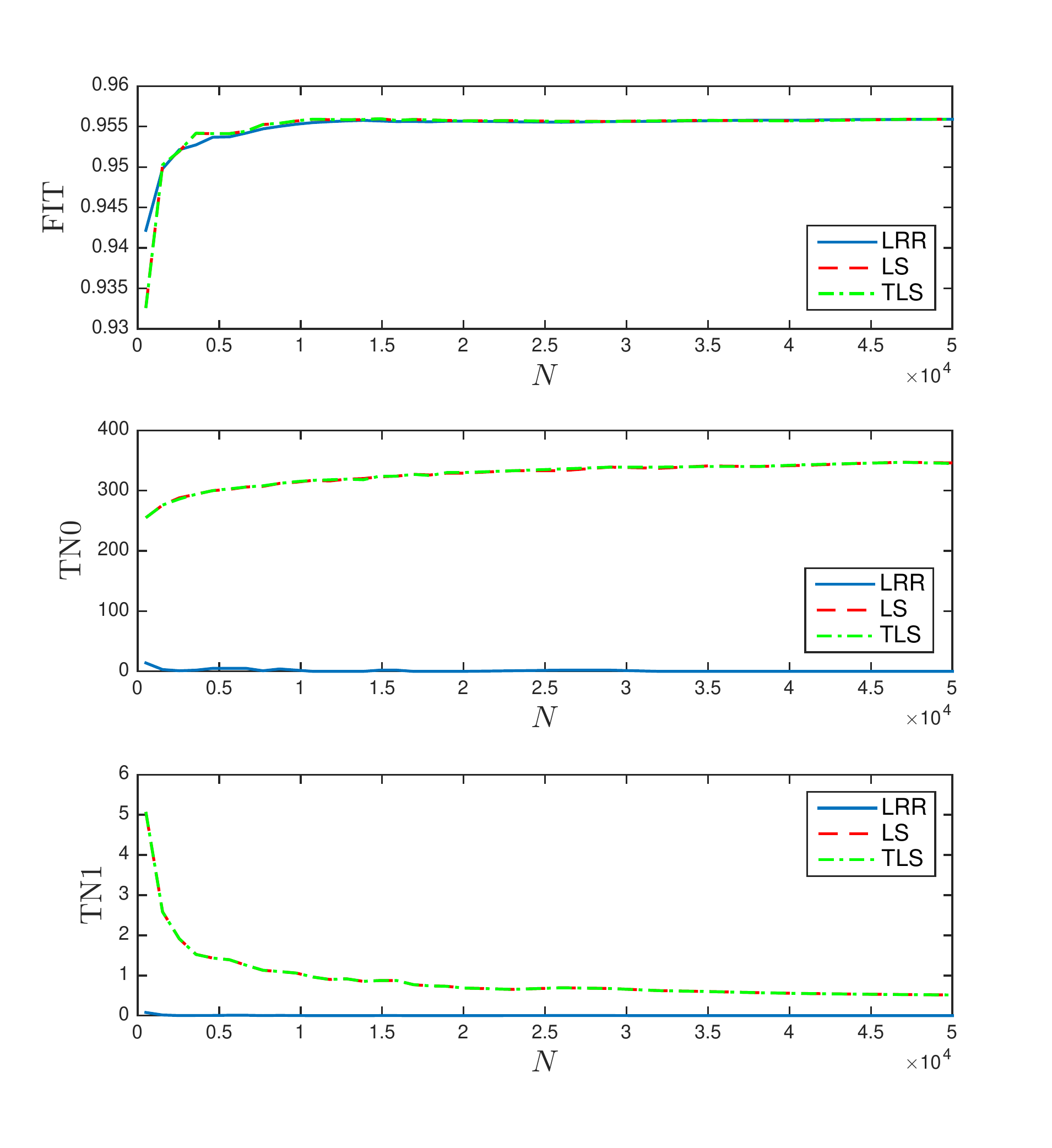}
\vspace*{-0.6cm}
\caption{Values of $\mbox{FIT}$, $\mbox{\mbox{\mbox{TN0}}}$ and $\mbox{TN1}$
for all models.}
\label{fig:conv}
\end{figure}

\begin{figure}[htb]
\vspace*{-0.0cm}\centering
\includegraphics[scale=0.45,bbllx=5mm,bblly=5mm,bburx=180mm,bbury=135mm,width=84mm,clip]{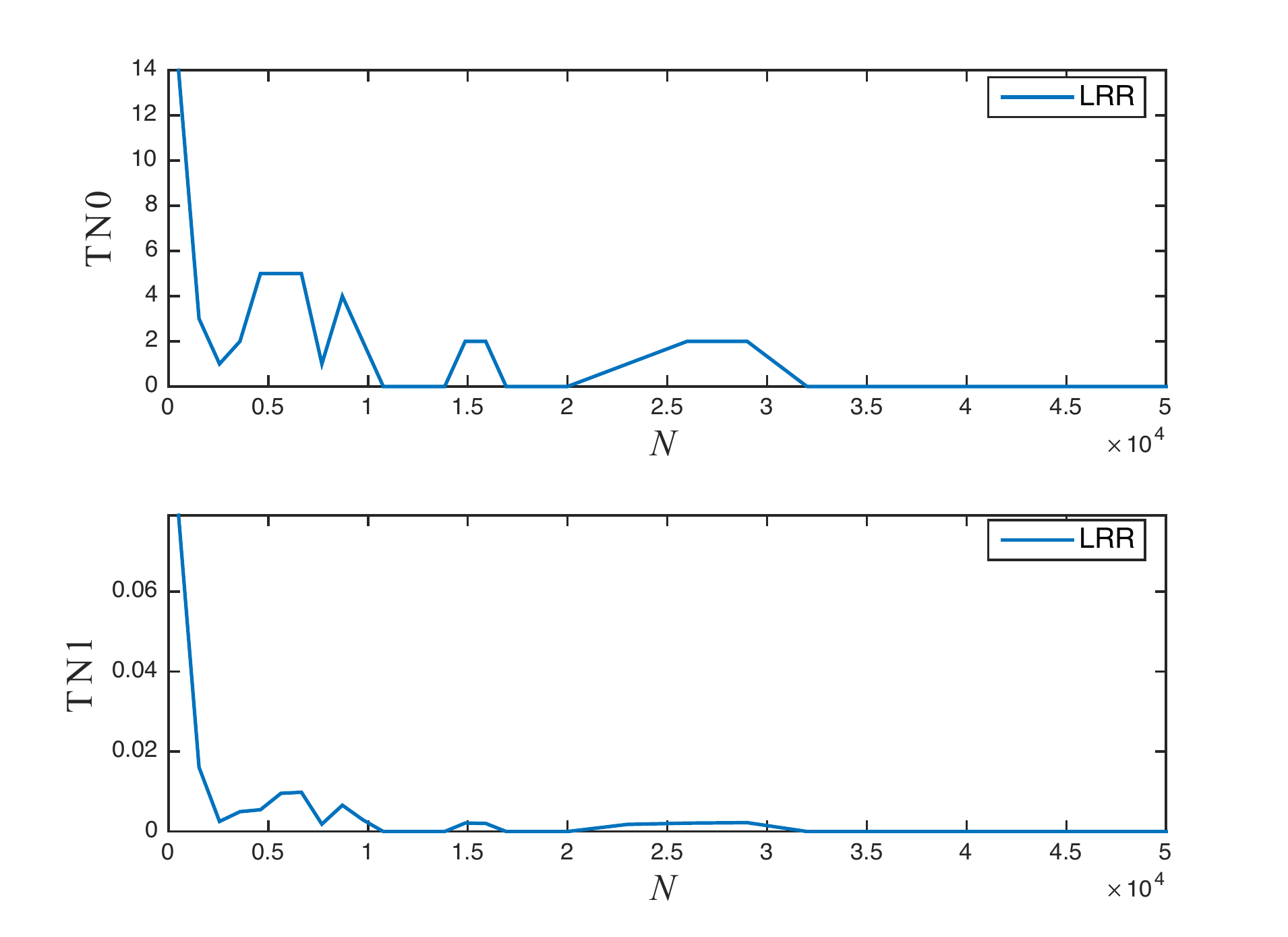}
\vspace*{-0.6cm}
\caption{Values of $\mbox{\mbox{\mbox{TN0}}}$ and $\mbox{TN1}$ for the
\mbox{LRR} models.}
\label{fig:conv2}
\end{figure}

A ``long'' i.i.d.\ input sequence with
zero mean and variance $\nu ^2=1$ was generated and corrupted by an i.i.d.\
noise with zero mean and variance $\sigma _u^2=0.03^2$. The true system was
then simulated using this input sequence, and the resulting output sequence
was corrupted by an i.i.d.\ noise with zero mean and variance $\sigma
_y^2=0.3^2$. The data corresponding to the output values with time index $%
k=1001,\ldots ,1000+N$ were selected, where $N=500,\ldots ,50000$. For each
value of $N$, an LLR, an LS and a TLS model were identified from these data.
The values of $\mbox{FIT}$, $\mbox{\mbox{\mbox{TN0}}}$ and $\mbox{TN1}$
obtained for these models are plotted as function of $N$ in Figures~\ref
{fig:conv} and \ref{fig:conv2}. We can observe that the three identification
methods lead to very similar $\mbox{FIT}$ values, the \mbox{LRR} models
giving slightly better results for low number of data. A key difference
between the three techniques is that the \mbox{LRR} method is able to select
the more appropriate impulse response components (i.e., the components with
index in the interval {[}1,~$n_l${]}), forcing the others to vanish. After a
certain value of $N$ (about 32000), the tail of the \mbox{LRR} models is
zero, confirming the theoretical result given in Corollary~\ref
{cor:main_lead}. Such an effective component selection is not guaranteed by
the other two methods which, on the contrary, have tails with support
cardinality (measured by the $\ell _0$ quasi-norm) that grows with $N$.

\subsection{Experimental data from a flexible robot arm}

\label{sec:example2}The identification of poorly damped systems from
experimental data is among the most challenging issues in many practical
applications. For this reason, as second test we considered a system with a
vibrating flexible robot arm described in \cite{TVSS98}, adopted as case
study in various software packages (\cite{KoPS94,Koll94,LabVIEW06}). Data
records from this process have been also analyzed in \cite{PiSc12,PDCD14}.
The input is the driving torque and the output is the tangential
acceleration of the tip of the robot arm. Ten consecutive periods of the
response to a multisine excitation signal were collected at a sampling
frequency of 500 Hz, for a total of 40960 data points.

We have built models using different techniques: the Leading Response
Recovery (LRR) and the regularized Diagonal/Correlated kernel (DCK) methods
to obtain high-order FIR models, the standard Prediction Error Method (PEM)
to estimate low-order state space models. Since the true system is unknown,
the models cannot be evaluated by their fit to the actual system. Instead,
we used the hold-out validation technique and measured how well the
identified models can reproduce the output on validation portions of the
data that were not used for estimation. We chose the estimation data to be
the portion 1:7000 and the validation data to be the portion 10000:40960.

\begin{figure}[tbh]
\vspace*{-0.0cm}\centering
\includegraphics[scale=0.45,width=84mm,clip]{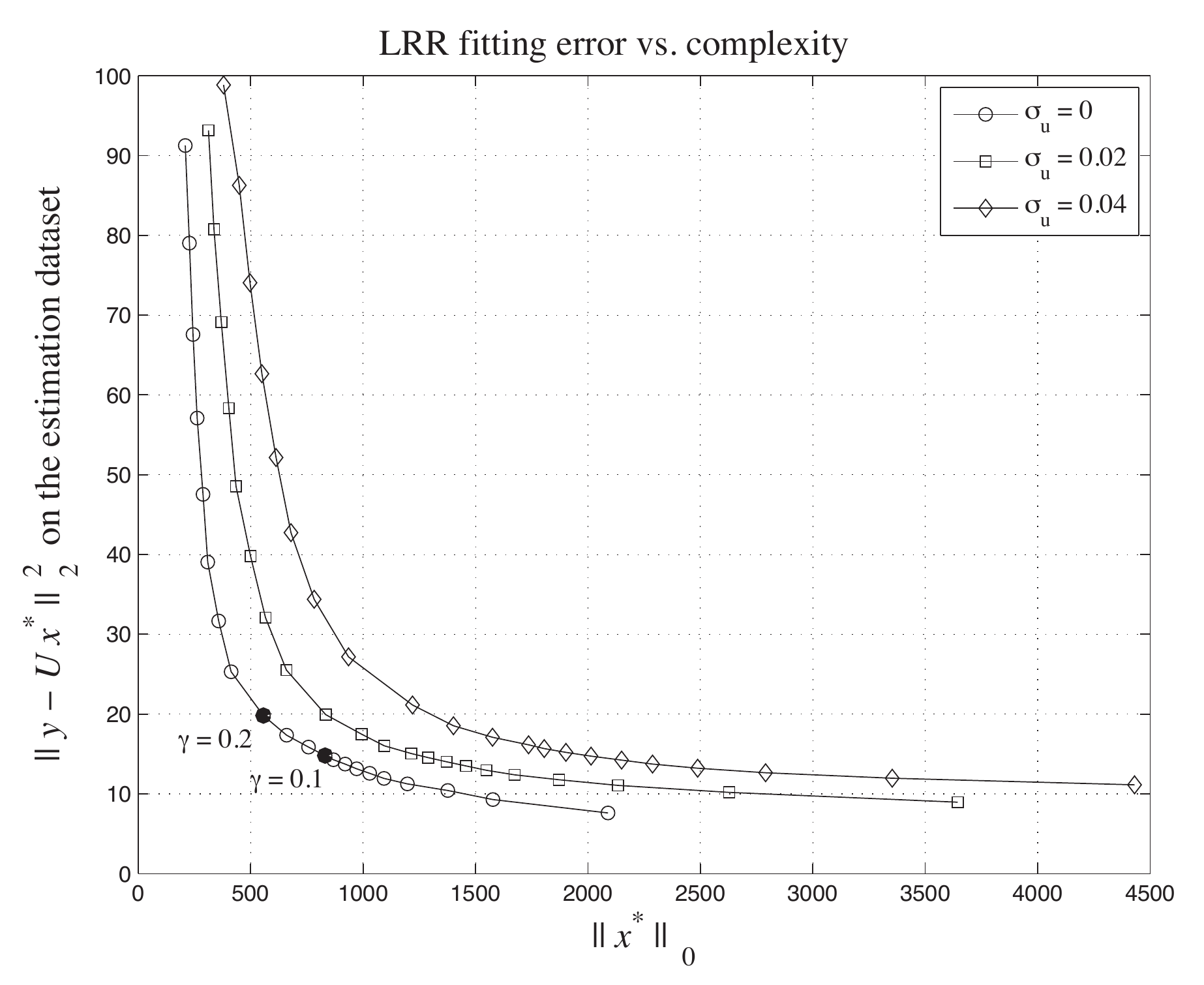} \vspace*{-0.6cm}
\caption{LRR fitting error $\left\| y-Ux^{*}\right\| _2^2$ vs.
complexity $\left\| x^{*}\right\| _0$.}
\label{fig:Pareto}
\end{figure}

To identify the LRR models, the procedure described in Section \ref
{sec:id_proc} has been applied to suitably choose the values of $\sigma _u$
and $\gamma $, taking {$q=7000$ as initial estimate length}. No a-priori
information was available about the input noise affecting the system, driven
by an input signal $u$ with sample variance $\nu ^2=0.0298$. For this
reason, three scenarios have been considered, with $\sigma _u=0$, $\sigma
_u=0.02$ and $\sigma _u=0.04$. These values correspond to noise-to-signal
ratios of $0\%$ (i.e., no-input-noise situation), $1.3\%$, and $5.3\%$,
respectively. Then the LRR algorithm has run with values of $\gamma $ in the
range $\left[ 0.01,\ 1\right] $, using the MATLAB's command \texttt{lasso}
with optional input arguments \texttt{'RelTol',4e-4,'Standardize',false}.
The results in terms of fitting error $\left\| y-Ux^{*}\right\| _2^2$ and
complexity $\left\| x^{*}\right\| _0$ are shown in Figure~\ref{fig:Pareto}
(lower error and higher complexity are achieved for lower values of $\gamma $%
). As expected, curves with lower $\sigma_u$ dominate curves with higher $%
\sigma_u$ (for any given $\gamma$, the solution obtained with lower $%
\sigma_u $ has both lower error and complexity with respect to a solution
obtained with higher $\sigma_u$). However, the choice of the actual curve to
use depends on our confidence on the true value of $\sigma_u$, and
underestimating this value may lead to worse-than-expected performance on
validation data. Also, curves with higher $\sigma_u$ show a flatter behavior
after the ``knee'' for lower $\gamma$ values.

We found a reasonable tradeoff for $\gamma =0.2$, allowing a satisfactory
fitting error (around $19.8$ for $\sigma _u=0$, which raises up to $21.1$ in
the worst-case $\sigma _u=0.04$) with a small complexity (around $560$, that
raises up to $1220$ for $\sigma _u=0.04$). Alternatively, $\gamma =0.1$
allows a lower error (around $14.8$ for $\sigma _u=0$, which raises up to $%
16.1$ for $\sigma _u=0.04$) with a still acceptable complexity (around $830$%
, that raises up to $1740$ for $\sigma _u=0.04$).

\begin{figure}[b]
\vspace*{-0.0cm}\centering
\includegraphics[scale=0.68,bbllx=20mm,bblly=70mm,bburx=195mm,bbury=215mm,width=83mm,clip]{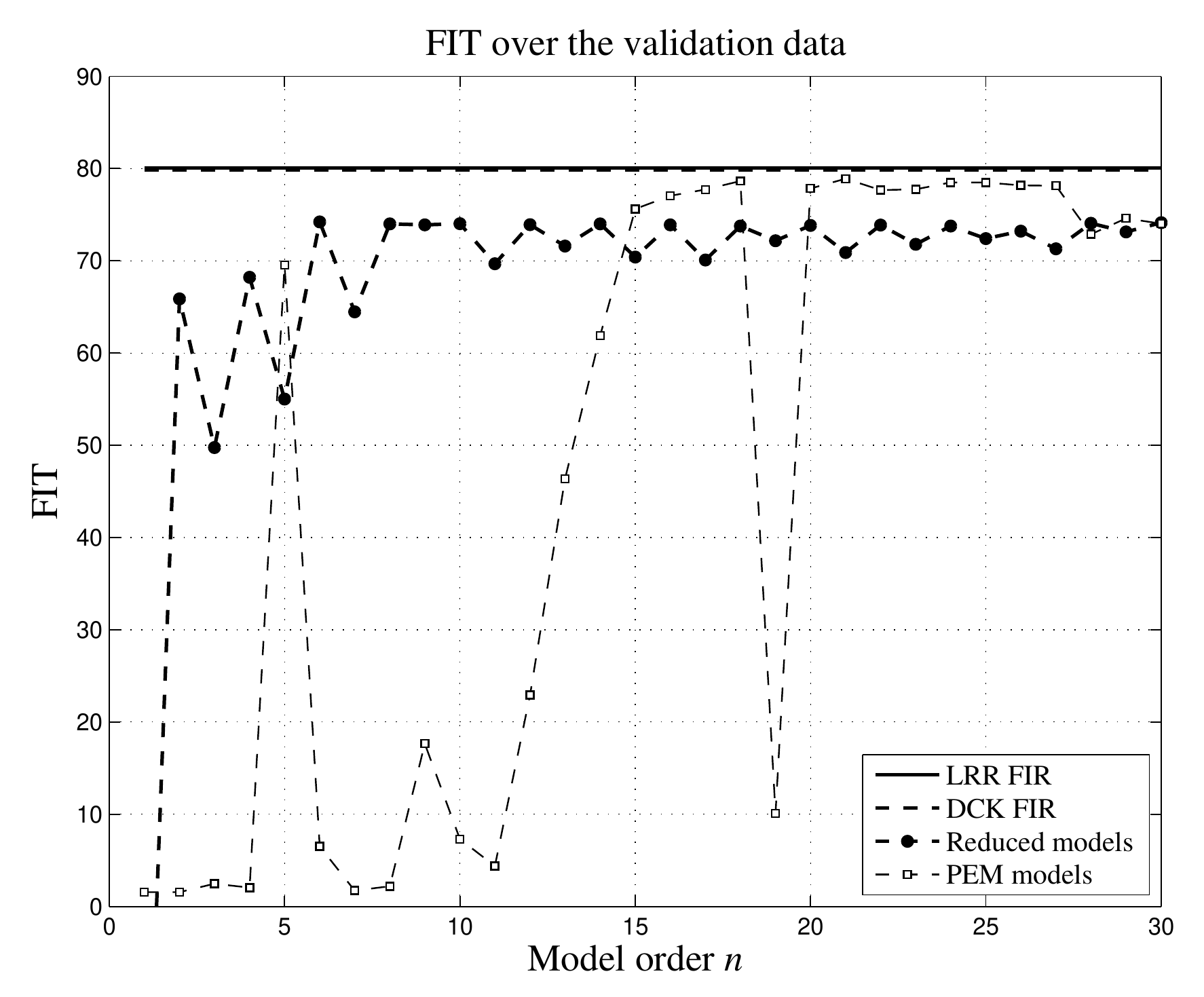}
\vspace*{-0.3cm}
\caption{Values of $\mbox{FIT}$ for all models.}
\label{fig:robot_arm_1}
\end{figure}

To fairly compare the performances achieved by LRR and DCK methods, FIR
models with the same given order have been identified. First, the LRR FIR
model of order $2500$ was identified with FIT value of 80.1\%. Then, the DCK
FIR model of order $2500$ was estimated using regularized least squares,
tuned by the marginalized likelihood method and with the unknown input data
set to zero, via the MATLAB's command \texttt{impulseest(data,2500,0,opt)}
with option \texttt{opt} set as \texttt{opt.RegulKernel='dc';} \texttt{opt.
Advanced.AROrder=0}. The FIT for this DCK FIR model was 79.9\%. For
illustration, the FIT values of the LRR and DCK FIR models are shown in
Figure~\ref{fig:robot_arm_1} as horizontal lines. For comparison, we
estimated $n$-th order state space PEM models without disturbance model for $%
n=1,\ldots ,30$ (via the MATLAB's command \texttt{pem(data,n,'dist','no')})
and calculated their FIT index to validation data. These FIT values are
shown as function of $n$ in Figure~\ref{fig:robot_arm_1}. The two best FITs
were 78.9\% and 78.6\%, obtained for order $n=21$ and $n=18$, respectively,
while the $5$-th order model with FIT value of 69.6\% could be a reasonable
tradeoff between accuracy and complexity. In any case, any PEM fit is worst
than those provided by the LRR and DCK FIR models.

One may observe that FIR models of order $2500$ are quite large, but it is
interesting to note that they can be easily reduced to low-order state space
models by model order reduction methods, like balanced truncation, Hankel
norm minimization and $\mathcal{L}_2$ reduction. For example, we applied the
square root balanced truncation method to the LRR FIR model, to obtain
reduced state space models of order $n=1,\ldots ,30$ (via the MATLAB's
command \texttt{balancmr}), and we computed their FIT index on validation
data. These FIT values are also shown as function of $n$ in Figure~\ref
{fig:robot_arm_1}. It can be observed that a reduced state-space model of
order $n=6$ provides a FIT of 74.2\%, which is better than any PEM-estimated
state space model of order $n=1,\ldots ,14$.

\begin{figure}[b]
\vspace*{-0.0cm}\centering
\includegraphics[scale=0.68,bbllx=20mm,bblly=70mm,bburx=195mm,bbury=215mm,width=83mm,clip]{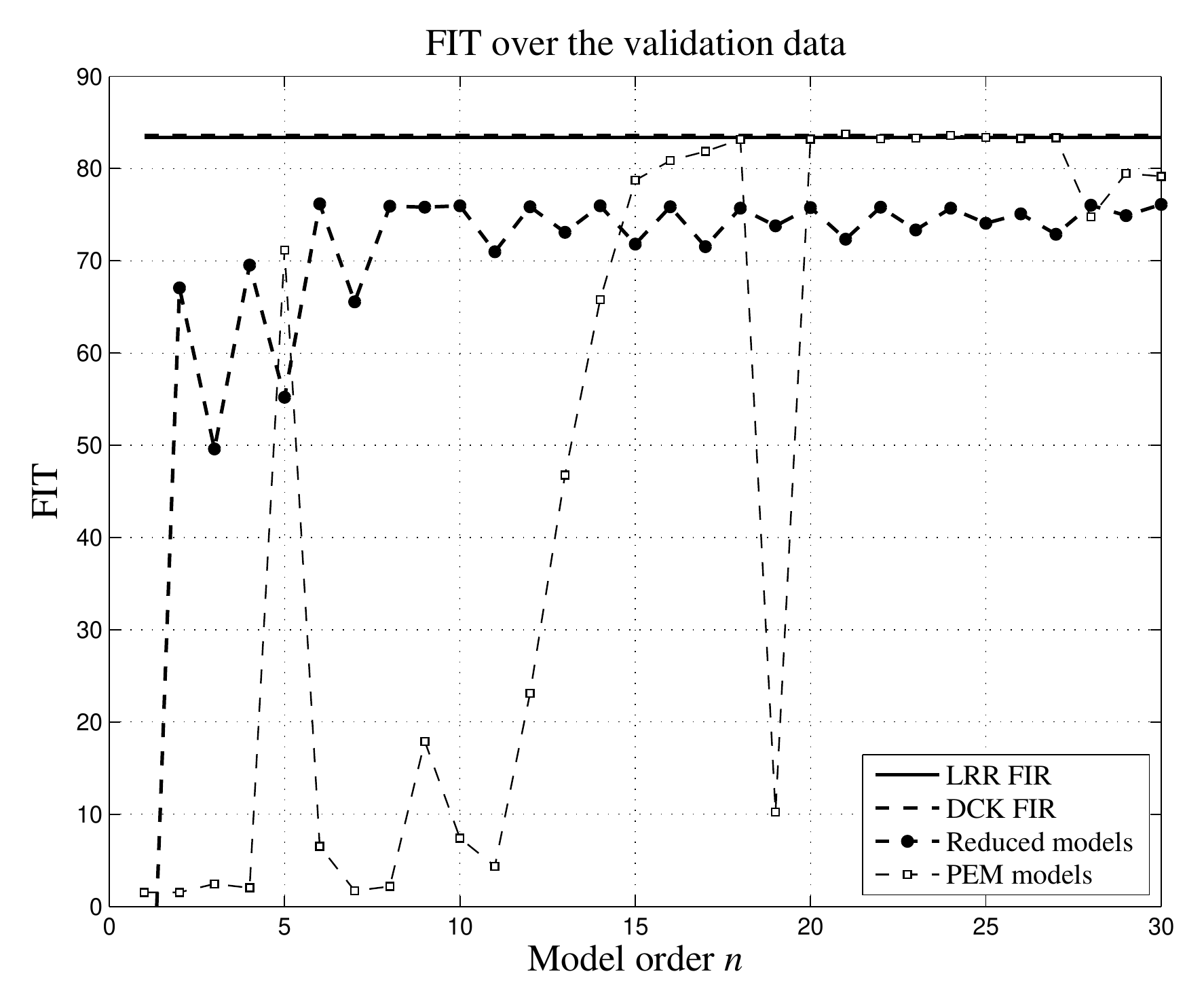}
\vspace*{-0.3cm}
\caption{Values of $\mbox{FIT}$, neglecting the starting 3000 samples.}
\label{fig:robot_arm_2}
\end{figure}

To discriminate the effects of the transient due to the mismatch between the
initial states of the actual system and the identified models, the FIT index
has been also computed by neglecting the initial 3000 samples of the
validation data. The FIT values of the LRR and DCK FIR models of order $2500$
raise to 83.4\% and 83.6\%, respectively; the FIT values of the PEM models
of order $n=5,18,21$ go to 71.2\%, 83.2\%, 83.7\%, respectively; the FIT
value of the reduced state-space model of order $n=6$ increases to 76.2\%.
All these results are shown in Figure~\ref{fig:robot_arm_2}.

\newpage

It is worth to observe that, even if the fit performances of LRR and DCK FIR
models are very close, the computational complexity of their corresponding
algorithms is dramatically different. Referring to a workstation equipped
with an Intel(R) Core(TM) i7-3770 CPU @ 3.40 GHz and with 16 GB of RAM, the
overall CPU time used to estimate the LRR FIR model of order $2500$ was
around $30$ seconds, while the computation of DCK FIR model of order $2500$
required around $8700$ seconds, i.e., $290$ times more. For comparison, FIR
models of order $3000$ and $3500$ were also identified using the same 7000
estimation data as before: the CPU times required by the LRR FIR models were
around $30$ and $36 $ seconds, respectively, while the CPU times required by
the DCK FIR models were around $11000$ and $23200$ seconds, respectively.
Widening the estimation data to the first 10000 samples, the identification
of LRR FIR models of orders up to 5000 required no more than $75$ seconds,
thus showing that the approach proposed in this paper scales nicely with the
problem dimensionality.

\section{Conclusions}

\label{sec:concl} A novel method for the identification of low-complexity
FIR models from experimental data is presented in this paper. The method is
based on an Elastic Net criterion, which considers an identification cost
defined as a weighted combination of a standard prediction error term, an $%
\ell _2$ regularization term, and a weighted $\ell _1$ penalty term. The
main novelty of the method with respect to the state of the art is that it
allows for an effective selection of the model order, while requiring only
stability and standard statistical assumptions on the noises affecting the
system; no additional information on the system impulse response behavior is
needed. The effectiveness of the method has been tested through both
extensive numerical simulations (considering two typical situations: one
with a fixed number of data, and one with an arbitrarily large number of
data) and real experimental data from a lightly damped mechanical system. In
all situations, the method showed high numerical efficiency and satisfactory
order selection capability and simulation accuracy.

Research activity is being devoted to developing a weighted version of the
method proposed here. It is indeed expected that including suitable weights
in the identification criterion may make the model order selection even more
efficient, especially in situations where a low number of data is available.


\TeXButton{small}{\small}\appendix

\section{Appendix}

\subsection{Proof of Lemma~\ref{lem:lip_conv}}

\label{sec:app_proof:lip_conv} From the hypothesis that $y_i\leadsto \bar{y}%
_i$, $i=1,\ldots ,p$, applying the definition of symbol $\leadsto $, we have
that for any $\tilde{\epsilon}>0$ and $\tilde{\beta}\in (0,1)$ there exists
an integer $\tilde{N}_{\tilde{\epsilon},\tilde{\beta}}$ such that

\noindent
\[
\mathbb{P}\{|y_i-\bar{y}_i|\leq \tilde{\epsilon}\}\geq 1-\tilde{\beta},\quad
\forall N\geq \tilde{N}_{\tilde{\epsilon},\tilde{\beta}}.
\]

\vspace*{-0.4em}

\noindent
From Bonferroni's inequality we further have that the probability of the
joint event $\{|y_i-\bar{y}_i|\leq \tilde{\epsilon},\,i=1,\ldots ,p\}$ is
lower bounded as

\noindent
\[
\mathbb{P}\{|y_i-\bar{y}_i|\leq \tilde{\epsilon},\,i=1,\ldots ,p\}\geq 1-p%
\tilde{\beta},\quad \forall N\geq \tilde{N}_{\tilde{\epsilon},\tilde{\beta}%
}.
\]

\vspace*{-0.4em}

\noindent
Since $\mathbb{P}\{|y_i-\bar{y}_i|\leq \tilde{\epsilon},\,i=1,\ldots ,p\}=%
\mathbb{P}\{\Vert y-\bar{y}\Vert _\infty \leq \tilde{\epsilon}\}$, letting $%
\beta \doteq p\tilde{\beta}$, we write

\noindent
\[
\mathbb{P}\{\Vert y-\bar{y}\Vert _\infty \leq \tilde{\epsilon}\}\geq 1-\beta
,\quad \forall N\geq \tilde{N}_{\tilde{\epsilon},\beta /p}.
\]

\vspace*{-0.4em}

\noindent
Now, from the hypothesis that $f$ is Lipschitz continuous, it follows that
there exists a finite constant $C\geq 0$ such that

\noindent
\[
|f(y)-f(\bar{y})|\leq C\Vert y-\bar{y}\Vert _\infty .
\]

\vspace*{-0.4em}

\noindent
Therefore, $\Vert y-\bar{y}\Vert _\infty \leq \tilde{\epsilon}$ implies that
$|f(y)-f(\bar{y})|\leq \epsilon $, for $\epsilon \doteq C\tilde{\epsilon}$,
whence

\noindent
\[
\mathbb{P}\{|f(y)-f(\bar{y})|\leq \epsilon \}\geq 1-\beta ,\quad \forall
N\geq \tilde{N}_{\epsilon /C,\beta /p},
\]

\vspace*{-0.4em}

\noindent
which proves that $f(y)\leadsto f(\bar{y})$. \qed

\subsection{Proof of Lemma~\ref{lem:recovery_1}}

\label{sec:app_proof:lem1} The claim is a direct consequence of the first
point of Theorem~8 in \cite{Tropp:06}, where the index set $\Lambda $ is $%
\{1,\ldots ,n\}$, and $\mbox{ERC}(\Lambda )$ in \cite{Tropp:06} coincides
with $\Upsilon _n(A)$. The symbol $a_\Lambda $ used in Theorem~8 of \cite
{Tropp:06} corresponds to $P_nb$, that is the best $\ell _2$ approximation
of $b$ using a linear combination of the first $n$ columns of $A$. These
first $n$ columns have unit $\ell _2$ norm and are indeed linearly
independent, as requested by the hypotheses of Theorem~8 in \cite{Tropp:06},
due to the specific structure of $A=\bar{A}T$, where $\bar{A}$, shown in (%
\ref{eq:bdef}), has a multiple of the identity matrix $I_q$ as a bottom
block. \qed

\subsection{Proof of Lemma~\ref{lem:convergence}}

\label{sec:app_proof:lem_convergence} Some parts of this result might
possibly be derived as a particular case of Theorem 2.3 in \cite
{Ljung:99_book}; we here report a full proof for the specific case of
interest in the present work. For $i=1,2,\ldots ,$ let us define $u_i=\left[
u(2\!-\!i)\;u(3\!-\!i)\;\cdots \;u(N\!+\!1\!-\!i)\right] ^{\top }\!\in {{%
\mathbb R}^N}$. Then, for all $i$ and $j$, we have that

\noindent
\[
\begin{array}{ll}
\frac 1Nu_i^{\top }\!u_j & =\frac 1N\left[ u(2\!-\!i)\,u(3\!-\!i)\cdots
u(N\!+\!1\!-\!i)\right] \cdot \\
& \hspace*{7.0mm}\left[ u(2\!-\!j)\,u(3\!-\!j)\cdots u(N\!+\!1\!-\!j)\right]
^{\top } \\
& =\frac 1N\sum_{k=2}^{N+1}u(k-i)u(k-j) \\
& =\frac 1N\sum_{k=1}^Nu(k+1-i)u(k+1-j).
\end{array}
\]

\vspace*{-0.4em}

\noindent
Consider first the case where $i=j$. Then

\noindent
\[
\tfrac 1Nu_i^{\top }u_i=\tfrac 1N\tsum\nolimits_{k=1}^Nu(k+1-i)^2
\]

\vspace*{-0.5em}

\noindent
is the empirical mean of the elements of the sequence of length $N$ of
random variables $x_k=u(k\!+\!1\!-\!i)^2$, $k=1,\ldots ,N$, such that, for
all $k$, $i$ and $l\neq k$:

\noindent
\[
\mathbb{E}\{x_k\}=\mathbb{E}\{u(k\!+\!1\!-\!i)^2\}={\mathrm{var}}%
\{u(k\!+\!1\!-\!i)\}=\nu ^2<\infty
\]

\vspace*{-1.5em}

\noindent
\begin{eqnarray*}
\hspace*{2mm}{\mathrm{var}}\{x_k\} &=&\mathbb{E}\left\{ \left( x_k-\mathbb{E}%
\left\{ x_k\right\} \right) ^2\right\} =\mathbb{E}\left\{ \left(
u(k\!+\!1\!-\!i)^2\!-\!\nu ^2\right) ^2\right\} \\
&=&\mathbb{E}\left\{ u(k\!+\!1\!-\!i)^4\!-\!2\nu
^2u(k\!+\!1\!-\!i)^2\!+\!\nu ^4\right\} \\
&=&\mathbb{E}\left\{ u(k\!+\!1\!-\!i)^4\right\} -2\nu ^2\mathbb{E}\left\{
u(k\!+\!1\!-\!i)^2\right\} +\nu ^4 \\
&=&\overline{m\TeXButton{rule}{\rule{0mm}{2mm}}_4}-\nu ^4<\infty
\end{eqnarray*}

\vspace*{-1.5em}

\noindent
\[
\begin{array}{l}
\mathbb{E}\left\{ \left( x_k\!-\!\mathbb{E}\left\{ x_k\right\} \right)
\left( x_l\!-\!\mathbb{E}\left\{ x_l\right\} \right) \right\} \\
\hspace*{2mm}=\mathbb{E}\left\{ \left( u(k\!+\!1\!-\!i)^2\!-\!\nu ^2\right)
\left( u(l\!+\!1\!-\!i)^2\!-\!\nu ^2\right) \right\} \\
\hspace*{2mm}=\mathbb{E}\left\{ u(k\!+\!1\!-\!i)^2u(l\!+\!1\!-\!i)^2+\right.
\\
\hspace*{2mm}\hspace*{8mm}\left. -\nu ^2\!\left[
u(k\!+\!1\!-\!i)^2\!+\!u(l\!+\!1\!-\!i)^2\right] \!+\!\nu ^4\right\} \\
\hspace*{2mm}=\mathbb{E}\left\{ u(k\!+\!1\!-\!i)^2u(l\!+\!1\!-\!i)^2\right\}
+ \\
\hspace*{2mm}\hspace*{3mm}-\nu ^2\!\left[ \mathbb{E}\left\{
u(k\!+\!1\!-\!i)^2\right\} \!+\!\mathbb{E}\left\{ u(l\!+\!1\!-\!i)^2\right\}
\right] \!+\!\nu ^4 \\
\hspace*{2mm}=\mathbb{E}\left\{ u(k\!+\!1\!-\!i)^2\right\} \mathbb{E}\left\{
u(l\!+\!1\!-\!i)^2\right\} \!-\!\nu ^4=\nu ^2\nu ^2-\nu ^4=0
\end{array}
\]

\vspace*{-0.8em}

\noindent
where the last derivation follows from the fact that $x_k$ and $x_l$ are
mutually independent since the input $u(k)$ is an i.i.d.\ sequence. By
applying the Chebyshev's inequality for sums of uncorrelated variables shown
in Section~\ref{sec:cheby}, it holds that

\noindent
\[
\tfrac 1Nu_i^{\top }u_i\leadsto \mathbb{E}\{x_k\}=\nu ^2,\quad \forall i.
\]

\vspace*{-0.5em}

\noindent
Consider next the case where $i\neq j$. Then

\noindent
\[
\tfrac 1Nu_i^{\top }\!u_j=\tfrac 1N\tsum\nolimits_{k=1}^Nu(k+1-i)u(k+1-j)
\]

\vspace*{-0.5em}

\noindent
is the empirical mean of the elements of the sequence of length $N$ of
random variables $x_k=u(k+1-i)u(k+1-j)$, $k=1,\ldots ,N$, such that, for all
$k$, $i$, $j=i+\tilde{i}\neq i$ and $l=k+\tilde{k}\neq k$, with $\tilde{i}%
\neq 0$ and $\tilde{k}\neq 0\,$:\vspace*{-1.2em}

\noindent
\begin{eqnarray*}
\hspace*{10mm}\mathbb{E}\{x_k\} &=&\mathbb{E}\left\{
u(k\!+\!1\!-\!i)u(k\!+\!1\!-\!j)\right\} \\
&=&\mathbb{E}\left\{ u(k\!+\!1\!-\!i)\right\} \mathbb{E}\left\{
u(k\!+\!1\!-\!j)\right\} =0
\end{eqnarray*}

\vspace*{-1em}

\noindent
\begin{eqnarray*}
{\mathrm{var}}\{x_k\} &=&\mathbb{E}\!\left\{ \!\left( x_k\!-\mathbb{E}%
\!\left\{ x_k\right\} \right) ^2\right\} \!=\!\mathbb{E}\!\left\{ \!\left(
u(k\!+\!1\!-\!i)u(k\!+\!1\!-\!j)\right) ^2\right\} \\
&=&\mathbb{E}\!\left\{ u(k\!+\!1\!-\!i)^2u(k\!+\!1\!-\!j)^2\right\} \\
&=&\mathbb{E}\!\left\{ u(k\!+\!1\!-\!i)^2\right\} \mathbb{E}\!\left\{
u(k\!+\!1\!-\!j)^2\right\} =\nu ^4<\infty
\end{eqnarray*}

\vspace*{-2.5em}

\noindent
\begin{eqnarray*}
\lefteqn{\mathbb{E}\!\left\{ \left( x_k\!-\!\mathbb{E}\!\left\{ x_k\right\}
\right) \left( x_l\!-\!\mathbb{E}\!\left\{ x_l\right\} \right) \right\} \!=}
\\
&=&\!\mathbb{E}\!\left\{ \left( u(k\!+\!1\!-\!i)u(k\!+\!1\!-\!j)\right)
\left( u(l\!+\!1\!-\!i)u(l\!+\!1\!-\!j)\right) \right\} \\
&=&\!\mathbb{E}\{u(\hspace*{-0.3mm}k\!+\!1\!-\!i)u(\hspace*{-0.3mm}%
k\!+\!1\!-\!i\!-\!\tilde{i})u(\hspace*{-0.3mm}k\!+\!\tilde{k}\!+\!1\!-\!i)u(%
\hspace*{-0.3mm}k\!+\!\tilde{k}\!+\!1\!-\!i\!-\!\tilde{i})\} \\
&=&\!\mathbb{E}\{u(\hspace*{-0.3mm}k\!+\!1\!-\!i)u(\hspace*{-0.3mm}%
k\!+\!1\!-\!i\!-\!\tilde{i})u(\hspace*{-0.3mm}k\!+\!1\!-\!i\!+\!\tilde{k})u(%
\hspace*{-0.3mm}k\!+\!1\!-\!i\!-\!\tilde{i}\!+\!\tilde{k})\}
\end{eqnarray*}

\vspace*{-1.2em}

\noindent
if $\tilde{i}=\tilde{k}$, then:\vspace*{-1.2em}

\noindent
\begin{eqnarray*}
\lefteqn{\mathbb{E}\left\{ \left( x_k\!-\!\mathbb{E}\left\{ x_k\right\}
\right) \left( x_l\!-\!\mathbb{E}\left\{ x_l\right\} \right) \right\} =} \\
&=&\mathbb{E}\left\{ u(k\!+\!1\!-\!i)^2u(k\!+\!1\!-\!i\!-\!\tilde{i}%
)u(k\!+\!1\!-\!i\!+\!\tilde{i})\right\} \\
&=&\mathbb{E}\left\{ u(k\!+\!1\!-\!i)^2\right\} \mathbb{E}\left\{
u(k\!+\!1\!-\!i\!-\!\tilde{i})\right\} \mathbb{E}\left\{ u(k\!+\!1\!-\!i\!+\!%
\tilde{i})\right\} =0
\end{eqnarray*}

\vspace*{-1.2em}

\noindent
otherwise, if $\tilde{i}\neq \tilde{k}$, then:\vspace*{-1.2em}

\noindent
\begin{eqnarray*}
\lefteqn{\mathbb{E}\left\{ \left( x_k\!-\!\mathbb{E}\left\{ x_k\right\}
\right) \left( x_l\!-\!\mathbb{E}\left\{ x_l\right\} \right) \right\} =} \\
&=&\mathbb{E}\{u(\hspace*{-0.3mm}k\!+\!1\!-\!i)u(\hspace*{-0.3mm}%
k\!+\!1\!-\!i\!-\!\tilde{i})u(\hspace*{-0.3mm}k\!+\!1\!-\!i\!+\!\tilde{k})u(%
\hspace*{-0.3mm}k\!+\!1\!-\!i\!-\!\tilde{i}\!+\!\tilde{k})\} \\
&=&\mathbb{E}\!\left\{ u(k\!+\!1\!-\!i)\right\} \cdot \\
&&\hspace*{0mm}\mathbb{E}\{u(k\!+\!1\!-\!i\!-\!\tilde{i})u(k\!+\!1\!-\!i\!+\!%
\tilde{k})u(k\!+\!1\!-\!i\!-\!\tilde{i}\!+\!\tilde{k})\}%
\mbox{\thinspace
=\thinspace 0}
\end{eqnarray*}

\vspace*{-1.2em}

\noindent
and this means that $x_k$ and $x_l$ are uncorrelated for all $k\neq l$. By
applying the Chebyshev's inequality for sums of uncorrelated variables shown
in Section~\ref{sec:cheby}, we obtain that

\noindent
\[
\tfrac 1Nu_i^{\top }u_j\leadsto \mathbb{E}\{x_k\}=0\text{, for all }i\neq j,
\]

\vspace*{-0.5em}

\noindent
which proves (\ref{lem:convergence_claima}).

\noindent We next prove (\ref{lem:convergence_claimb}). Since $\delta
_y\!=\!\left[ \delta _y(1)\,\delta _y(2)\,\cdots \,\delta _y(N)\right]
^{\top }\!\!\in \!{{\mathbb R}^N\!}$, then for $i=1,2,\ldots $:%
\vspace*{-1.2em}

\noindent
\begin{eqnarray*}
\hspace*{10mm}\tfrac 1Nu_i^{\top }\!\delta _y &=&\tfrac 1N\left[
u(2\!-\!i)\,u(3\!-\!i)\cdots u(N\!+\!1\!-\!i)\right] \cdot \\
&&\hspace*{3.6mm}\left[ \delta _y(1)\;\delta _y(2)\;\cdots \;\delta
_y(N)\right] ^{\top } \\
&=&\tfrac 1N\tsum\nolimits_{k=1}^Nu(k+1-i)\delta _y(k)
\end{eqnarray*}

\vspace*{-1.2em}

\noindent is the empirical mean of the elements of the sequence of length $N$
of random variables $x_k=u(k+1-i)\delta _y(k)$, $k=1,\ldots ,N$, such that,
for all $k$, $i$ and $l\neq k$:

\noindent
\[
\mathbb{E}\{x_k\}\!=\!\mathbb{E}\{u(k+1-i)\delta _y(k)\}\!=\!\mathbb{E}%
\{u(k+1-i)\}\mathbb{E}\{\delta _y(k)\}\!=\!0
\]

\vspace*{-2.3em}

\noindent
\begin{eqnarray*}
\hspace*{2mm}{\mathrm{var}}\{x_k\}\! &=&\!\mathbb{E}\left\{ \left( x_k-%
\mathbb{E}\left\{ x_k\right\} \right) ^2\right\} \\
\! &=&\!\mathbb{E}\left\{ \left( u(k\!+\!1\!-\!i)\delta _y(k)\right)
^2\right\} \!=\!\mathbb{E}\left\{ u(k\!+\!1\!-\!i)^2\delta _y(k)^2\right\} \\
\! &=&\!\mathbb{E}\left\{ u(k\!+\!1\!-\!i)^2\right\} \mathbb{E}\left\{
\delta _y(k)^2\right\} =\nu ^2\sigma _y^2<\infty
\end{eqnarray*}

\vspace*{-2.5em}

\noindent
\begin{eqnarray*}
\lefteqn{\mathbb{E}\left\{ \left( x_k\!-\!\mathbb{E}\left\{ x_k\right\}
\right) \left( x_l\!-\!\mathbb{E}\left\{ x_l\right\} \right) \right\} =} \\
\hspace*{2mm} &=&\mathbb{E}\left\{ \left( u(k\!+\!1\!-\!i)\delta
_y(k)\right) \left( u(l\!+\!1\!-\!i)\delta _y(l)\right) \right\} \\
\hspace*{2mm} &=&\mathbb{E}\left\{ u(k\!+\!1\!-\!i)u(l\!+\!1\!-\!i)\delta
_y(k)\delta _y(l)\right\} \\
\hspace*{2mm} &=&\mathbb{E}\left\{ u(k\!+\!1\!-\!i)\right\} \mathbb{E}%
\left\{ u(l\!+\!1\!-\!i)\right\} \mathbb{E}\left\{ \delta _y(k)\right\} %
\mathbb{E}\left\{ \delta _y(l)\right\} =0\text{.}
\end{eqnarray*}

\vspace*{-1.2em}

\noindent By applying the Chebyshev's inequality for sums of uncorrelated
variables shown in Section~\ref{sec:cheby}, it thus holds that

\noindent
\[
\tfrac 1Nu_i^{\top }\delta _y\leadsto \mathbb{E}\{x_k\}=0\text{, for all }i%
\text{.}
\]

\vspace*{-0.5em}

\noindent Finally, we prove (\ref{lem:convergence_claimc}). For $%
j=1,2,\ldots ,$ let us define $\delta _j=\left[ \delta _u(2\!-\!j)\;\delta
_u(3\!-\!j)\;\cdots \;\delta _u(N\!+\!1\!-\!j)\right] ^{\top }\!\in {{%
\mathbb R}^N}$. Then, $\forall i,j$:\vspace*{-1.2em}

\noindent
\begin{eqnarray*}
\hspace*{8mm}\tfrac 1Nu_i^{\top }\!\delta _j &=&\tfrac 1N\left[
u(2\!-\!i)\,u(3\!-\!i)\cdots u(N\!+\!1\!-\!i)\right] \cdot \\
&&\hspace*{3.6mm}\left[ \delta _u(2\!-\!j)\,\delta _u(3\!-\!j)\cdots \delta
_u(N\!+\!1\!-\!j)\right] ^{\top } \\
&=&\tfrac 1N\tsum\nolimits_{k=2}^{N+1}u(k-i)\delta _u(k-j) \\
&=&\tfrac 1N\tsum\nolimits_{k=1}^Nu(k+1-i)\delta _u(k+1-j)
\end{eqnarray*}

\vspace*{-1.2em}

\noindent is the empirical mean of the elements of the sequence of length $N$
of random variables $x_k=u(k+1-i)\delta _u(k+1-j)$, $k=1,\ldots ,N$, such
that, for all $k$, $i$, $j$ and $l\neq k$:\vspace*{-1.2em}

\noindent
\begin{eqnarray*}
\hspace*{8mm}\mathbb{E}\{x_k\} &=&\mathbb{E}\left\{ u(k\!+\!1\!-\!i)\delta
_u(k\!+\!1\!-\!j)\right\} \\
&=&\mathbb{E}\left\{ u(k\!+\!1\!-\!i)\right\} \mathbb{E}\left\{ \delta
_u(k\!+\!1\!-\!j)\right\} =0
\end{eqnarray*}

\vspace*{-2.5em}

\noindent
\begin{eqnarray*}
{\mathrm{var}}\{x_k\}\! &=&\!\mathbb{E}\!\left\{ \left( x_k\!-\mathbb{E}%
\!\left\{ x_k\right\} \right) ^2\right\} \!=\!\mathbb{E}\!\left\{ \!\left(
u(k\!+\!1\!-\!i)\delta _u(k\!+\!1\!-\!j)\right) ^2\right\} \\
\! &=&\!\mathbb{E}\!\left\{ u(k\!+\!1\!-\!i)^2\delta
_u(k\!+\!1\!-\!j)^2\right\} \! \\
\! &=&\!\mathbb{E}\!\left\{ u(k\!+\!1\!-\!i)^2\right\} \mathbb{E}\!\left\{
\delta _u(k\!+\!1\!-\!j)^2\right\} =\nu ^2\sigma _u^2<\infty
\end{eqnarray*}

\vspace*{-2.3em}

\noindent
\begin{eqnarray*}
\lefteqn{\mathbb{E}\left\{ \left( x_k\!-\!\mathbb{E}\left\{ x_k\right\}
\right) \left( x_l\!-\!\mathbb{E}\left\{ x_l\right\} \right) \right\} =} \\
\hspace*{3mm} &=&\mathbb{E}\left\{ \left( u(k\!+\!1\!-\!i)\delta
_u(k\!+\!1\!-\!j)\right) \left( u(l\!+\!1\!-\!i)\delta
_u(l\!+\!1\!-\!j)\right) \right\} \\
\hspace*{3mm} &=&\mathbb{E}\left\{ u(k\!+\!1\!-\!i)u(l\!+\!1\!-\!i)\delta
_u(k\!+\!1\!-\!j)\delta _u(l\!+\!1\!-\!j)\right\} \\
\hspace*{3mm} &=&\mathbb{E}\left\{ u(k\!+\!1\!-\!i)\right\} \mathbb{E}%
\left\{ u(l\!+\!1\!-\!i)\right\} \cdot \\
&&\hspace*{3mm}\mathbb{E}\left\{ \delta _u(k\!+\!1\!-\!j)\right\} \mathbb{E}%
\left\{ \delta _u(l\!+\!1\!-\!j)\right\} \mbox{$=0\text{.}$}
\end{eqnarray*}

\vspace*{-1.2em}

\noindent By applying the Chebyshev's inequality for sums of uncorrelated
variables shown in Section~\ref{sec:cheby}, it holds that\vspace*{-1.0em}

\[
\tfrac 1Nu_i^{\top }\delta _j\leadsto \mathbb{E}\{x_k\}=0\text{, for all }i%
\text{ and }j\text{.}
\]

\vspace*{-1.8em}\hspace*{80mm}\qed

\subsection{Proof of Theorem~\ref{thm:main}}

\label{sec:app_proof:thm_main}

\subsubsection{Preliminaries}

\label{sec:app_proof:thm_main_preliminaries} For any integer $n\leq q$, let $%
h^n\doteq [h(1)\,\cdots \,h(n)\,0\,\cdots \,0]^{\top }\in {\ {\mathbb R}^q}$
denote the $n$-leading truncation of $h_{\uparrow q}\in {\ {\mathbb R}^q}$,
let

\noindent
\[
h_{\downarrow n}=[h(n+1)\,h(n+2)\,\cdots ]^{\top },
\]

\vspace*{-0.5em}

\noindent and let, for $i=1,2,\ldots ,$

\noindent
\[
\tilde{u}_i\!\doteq \!\!\left[ \!\!
\begin{array}{c}
\tilde{u}(2-i)\vspace*{-1.0mm} \\
\tilde{u}(3-i)\vspace*{-2.0mm} \\
\vdots \vspace*{-1.5mm} \\
\tilde{u}(N\!+\!1-i)
\end{array}
\!\!\right] \!\!,\,u_i\!\doteq \!\!\left[ \!\!
\begin{array}{c}
u(2-i)\vspace*{-1.0mm} \\
u(3-i)\vspace*{-2.0mm} \\
\vdots \vspace*{-1.5mm} \\
u(N\!+\!1-i)
\end{array}
\!\!\right] \!\!,\,\delta _i\!\doteq \!\!\left[ \!\!
\begin{array}{c}
\delta _u(2-i)\vspace*{-1.0mm} \\
\delta _u(3-i)\vspace*{-2.0mm} \\
\vdots \vspace*{-1.5mm} \\
\delta _u(N\!+\!1-i)
\end{array}
\!\!\right] \!\!.
\]

\noindent For any integer $k\geq 1$, let $\Delta _{\uparrow k}\doteq [\delta
_1\,\cdots \,\delta _k]$, $\Delta _{\downarrow k}\doteq [\delta
_{k+1}\,\cdots ]$ and define $\Delta _{\downarrow 0}=[\delta _1\,\delta
_2\,\cdots ]\in {{\mathbb R}^{N,\infty }}$. Considering the expression in (%
\ref{eq:yk_inf_vec}), and splitting the summation at $n$, we can write

\noindent
\[
y=\tilde{U}_{\uparrow n}h_{\uparrow n}+\delta _y+\tilde{U}_{\downarrow
n}h_{\downarrow n}.
\]

\vspace*{-0.5em}

\noindent Further, using (\ref{eq_input_noise}), we have

\noindent
\[
y=U_{\uparrow n}h_{\uparrow n}+(\Delta _{\uparrow n}h_{\uparrow n}+\delta
_y+U_{\downarrow n}h_{\downarrow n}+\Delta _{\downarrow n}h_{\downarrow n}).
\]

\vspace*{-0.5em}

\noindent Since $U_{\uparrow n}h_{\uparrow n}=U_{\uparrow q}h^n\doteq Uh^n$,
we can write

\noindent
\[
y=Uh^n+e_0,\label{eq:ypluse}
\]

\vspace*{-0.5em}

\noindent where

\noindent
\[
e_0\doteq U_{\downarrow n}h_{\downarrow n}+\Delta _{\downarrow 0}h+\delta _y,%
\label{eq:ezero}
\]

\vspace*{-0.5em}

\noindent being $h\doteq [h(1)\,h(2)\,\cdots ]^{\top }$. Then, using the
notation in (\ref{eq:bdef}), we have that

\noindent
\[
b=\bar{A}h^n+e,
\]

\vspace*{-0.8em}

\noindent where

\noindent
\begin{equation}
e\doteq \left[
\begin{array}{c}
\vspace*{-7.0mm} \\
e_0\vspace*{-1.0mm} \\
-\sigma _u\sqrt{N}h^n
\end{array}
\right] ,  \label{eq:eeq}
\end{equation}

\vspace*{-0.8em}

\noindent and, by the change of variable $\tilde{h}^n=T^{-1}h^n$,

\noindent
\[
b=A\tilde{h}^n+e.\label{eq:beq}
\]

\vspace*{-0.5em}

\noindent Since $A=\bar{A}T$, where $T$ is diagonal, we can write

\noindent
\[
A_{\uparrow n}=\bar{A}_{\uparrow n}T_{\sharp n},
\]

\vspace*{-0.5em}

\noindent where $T_{\sharp n}$ is the $n\times n$ principal submatrix of $T$%
. Therefore,\vspace*{-1.2em}

\noindent
\begin{eqnarray*}
\hspace*{5mm}A_{\uparrow n}^{\dagger } &=&(A_{\uparrow n}^{\top }A_{\uparrow
n})^{-1}A_{\uparrow n}^{\top }=T_{\sharp n}^{-1}\bar{A}_{\uparrow
n}^{\dagger } \\
&=&T_{\sharp n}^{-1}\left( U_{\uparrow n}^{\top }U_{\uparrow n}+N\sigma
_u^2I_n\right) ^{-1}\left[
\begin{array}{cc}
U_{\uparrow n}^{\top } & \sigma _u\sqrt{N}I_{\uparrow n}^{\top }
\end{array}
\right] ,
\end{eqnarray*}

\vspace*{-1.2em}

\noindent where $I_{\uparrow n}$ is the submatrix formed by the first $n$
columns of the identity matrix $I_q$.

\noindent The orthogonal projector $P_n$ onto the span of $A_{\uparrow n}$
is given by\vspace*{-1.2em}

\noindent
\begin{eqnarray*}
P_n &=&A_{\uparrow n}A_{\uparrow n}^{\dagger }=\bar{A}_{\uparrow n}\bar{A}%
_{\uparrow n}^{\dagger } \\
&=&\left[ \!
\begin{array}{c}
\vspace*{-7.0mm} \\
U_{\uparrow n}\vspace*{-1.0mm} \\
\sigma _u\sqrt{N}I_{\uparrow n}
\end{array}
\!\right] \!\left( U_{\uparrow n}^{\top }U_{\uparrow n}\!+\!N\sigma
_u^2I_n\right) ^{-1}\!\left[ \!
\begin{array}{cc}
U_{\uparrow n}^{\top } & \sigma _u\sqrt{N}I_{\uparrow n}^{\top }
\end{array}
\!\right] .
\end{eqnarray*}

\vspace*{-1.2em}

\noindent For any given vector $b$, the best $\ell _2$ approximation of $b$
using the columns in $A_{\uparrow n}$ is given by $b_n=P_nb$, where, by the
Projection theorem, $b_n\bot (b-b_n)$. The corresponding optimal coefficient
vector is $x_n=A_{\uparrow n}^{\dagger }b=A_{\uparrow n}^{\dagger }b_n$.

\noindent For a column $a_i$ of $A$, $i=1,\ldots ,q$, we have that%
\vspace*{-1.2em}

\noindent
\begin{eqnarray*}
A_{\uparrow n}^{\dagger }a_i &=&t_iA_{\uparrow n}^{\dagger }\bar{a}_i \\
&=&t_iT_{\sharp n}^{-1}(U_{\uparrow n}^{\top }U_{\uparrow n}/N\!+\sigma
_u^2I_n)^{-1}(U_{\uparrow n}^{\top }u_i/N\!+\sigma _u^2I_{\uparrow n}^{\top
}\zeta _i),
\end{eqnarray*}

\vspace*{-1.2em}

where $\zeta _i$ is the $i$-th column of the identity matrix $I_q$, and $%
t_i\doteq [T]_{i,i}=\Vert \bar{a}_i\Vert _2^{-1}=\left( u_i^{\top
}u_i+N\sigma _u^2\right) ^{-1/2}$.

\noindent We shall next examine the condition in (\ref{eq:lem_condition}).

\subsubsection{The large $N$ sparsity pattern}

\label{sec:app_proof:thm_main_large_N} From Lemma~\ref{lem:convergence}, we
have that $U_{\uparrow n}^{\top }U_{\uparrow n}/N\!\leadsto \!\nu ^2I_n$,
and $U_{\uparrow n}^{\top }u_i/N\!\leadsto 0_{n,1}$, if $i>n$. Moreover, $%
t_iT_{\sharp n}^{-1}\leadsto I_n$, and $I_{\uparrow n}^{\top }\zeta
_i\!=\!\left[ \!
\begin{array}{cc}
I_n & 0_{n,q-n}
\end{array}
\!\right] \zeta _i=0_{n,1}$, if $i>n$. Therefore, considering the
scalar-valued function $\Vert W_{\sharp n}A_{\uparrow n}^{\dagger }a_i\Vert
_1$, which is Lipschitz continuous w.r.t.\ the entries of $U_{\uparrow
n}^{\top }U_{\uparrow n}/N$ and $U_{\uparrow n}^{\top }u_i/N$, and applying
Lemma~\ref{lem:lip_conv}, we obtain that, for $i>n$,\vspace*{-0.0em}

\noindent
\[
\begin{array}{l}
\left\| W_{\sharp n}A_{\uparrow n}^{\dagger }a_i\right\| _1\TeXButton{rule}
{\rule[-3mm]{0mm}{3mm}}\leadsto \\
\left\| W_{\sharp n}t_iT_{\sharp n}^{-1}\tfrac{\sigma _u^2}{\nu ^2+\sigma
_u^2}I_{\uparrow n}^{\top }\zeta _i\right\| _1=\left\| W_{\sharp n}I_n\tfrac{%
\sigma _u^2}{\nu ^2+\sigma _u^2}0_{n,1}\right\| _1=0.\vspace*{0em}
\end{array}
\]

\vspace*{-0.7em}

\noindent Hence it holds that\vspace*{-0.0em}

\noindent
\begin{equation}
\Upsilon _n(A)=1-\max_{i>n}w_i^{-1}\left\| W_{\sharp n}A_{\uparrow
n}^{\dagger }a_i\right\| _1\leadsto 1-0=1.  \label{eq:upsilon_n}
\end{equation}

\vspace*{-0.7em}

\noindent Let us now consider the left-hand side in the condition (\ref
{eq:lem_condition}). Using the fact that $b=\bar{A}h^n+e$, with $e$ given in
(\ref{eq:eeq}), we have

\noindent
\begin{equation}
\begin{array}{l}
W^{-1}A^{\top }(b-\!P_nb)=W^{-1}T\bar{A}^{\top }(\bar{A}h^n+e-\!P_n\bar{A}%
h^n\!-\!P_ne)\text{{}} \\
=W^{-1}T(\bar{A}^{\top }\!\bar{A}h^n\!+\bar{A}^{\top }e-\bar{A}^{\top }\!P_n%
\bar{A}h^n\!-\bar{A}^{\top }\!P_ne).\vspace*{-1.3em}
\end{array}
\label{eq:AbPb0}
\end{equation}

\vspace*{0em}

\noindent Defining $\tilde{T}\doteq T\sqrt{N}$ and dividing (\ref{eq:AbPb0})
by $\sqrt{N}$, we obtain\vspace*{-0.0em}

\noindent
\begin{equation}
\begin{array}{l}
\frac 1{\sqrt{N}}W^{-1}A^{\top }(b-P_nb)= \\
=W^{-1}\tilde{T}(\bar{A}^{\top }\!\bar{A}h^n\!+\!\bar{A}^{\top }\!e-\!\bar{A}%
^{\top }\!P_n\bar{A}h^n\!-\!\bar{A}^{\top }\!P_ne)\left/ N\right. \!.%
\vspace*{1.8em}
\end{array}
\label{eq:AbPb}
\end{equation}

\vspace*{-2.5em}

\noindent Now we evaluate\vspace*{-1.0em}

\noindent
\begin{eqnarray*}
\hspace*{4mm}\bar{A}^{\top }\!\bar{A}/\!N\! &=&U^{\top }\!U/\!N+\sigma
_u^2I_q \\
\bar{A}^{\top }\!e/\!N\! &=&U^{\top }\!U_{\downarrow n}h_{\downarrow
n}/\!N\!+U^{\top }\!\Delta _{\downarrow 0}h/\!N\!+U^{\top }\!\delta
_y/\!N\!-\!\sigma _u^2h^n \\
\bar{A}^{\top }\!\!P_n\bar{A}h^n\!/\!N\! &=&(U^{\top }\!U/\!N+\sigma
_u^2I_q)h^n \\
\bar{A}^{\top }\!\!P_ne/\!N\! &=&(U^{\top }\!U_{\uparrow n}/\!N+\sigma
_u^2I_{\uparrow n})(U_{\uparrow n}^{\top }U_{\uparrow n}/\!N+\sigma
_u^2I_n)^{-1}\cdot \\
&&\{[U_{\uparrow n}^{\top }U_{\downarrow n}h_{\downarrow
n}\!\!+\!U_{\uparrow n}^{\top }(\Delta _{\downarrow 0}h\!+\!\delta
_y)]/\!N\!-\!\sigma _u^2I_{\uparrow n}^{\top }h^n\}
\end{eqnarray*}

\vspace*{-1.0em}

\noindent and observe that\vspace*{-1.2em}

\noindent
\begin{eqnarray*}
\hspace*{25mm}U^{\top }U/N &\leadsto &\nu ^2I_q \\
U^{\top }U_{\downarrow n}h_{\downarrow n}/N &\leadsto &\nu ^2(h^q-h^n) \\
U^{\top }U_{\uparrow n}/N &\leadsto &\left[
\begin{array}{c}
\vspace*{-6.0mm} \\
\nu ^2I_n\vspace*{-1.0mm} \\
0\vspace*{-1.0mm}
\end{array}
\right]
\end{eqnarray*}
\begin{eqnarray*}
U^{\top }\Delta _{\downarrow 0}h/N &\leadsto &0 \\
\hspace*{25mm}U^{\top }\delta _y/N &\leadsto &0 \\
U_{\uparrow n}^{\top }U_{\downarrow n}h_{\downarrow n}/N &\leadsto &0
\end{eqnarray*}

\vspace*{-1.2em}

\noindent hence\vspace*{-1.0em}

\noindent
\begin{eqnarray*}
\hspace*{25mm}\bar{A}^{\top }\bar{A}/N &\leadsto &(\nu ^2+\sigma _u^2)I_q \\
\bar{A}^{\top }e/N &\leadsto &\nu ^2(h^q-h^n)-\sigma _u^2h^n \\
\bar{A}^{\top }P_n\bar{A}h^n/N &\leadsto &(\nu ^2+\sigma _u^2)h^n \\
\bar{A}^{\top }P_ne/N &\leadsto &-\sigma _u^2h^n.
\end{eqnarray*}

\vspace*{-1.2em}

\noindent Substituting in (\ref{eq:AbPb}) we obtain that

\noindent
\begin{equation}
\tfrac 1{\sqrt{N}}W^{-1}A^{\top }(b-P_nb)\leadsto {\nu ^2}W^{-1}\tilde{T}%
(h^q-h^n).  \label{eq:prf:lhs}
\end{equation}

\vspace*{-0.5em}

\noindent Finally, observe that for the $i$-th diagonal element $t_i$ of $T$
it holds that (by Lemma~\ref{lem:lip_conv})

\noindent
\[
t_i^2=\frac 1{\Vert \bar{a}_i\Vert _2^2}=\frac 1{\Vert u_i\Vert _2^2+\sigma
_u^2N}\leadsto \frac 1{N(\nu ^2+\sigma _u^2)}
\]

\vspace*{-0.5em}

\noindent and thus, for the $i$-th diagonal element $\tilde{t}_i$ of $\tilde{%
T}$, we have

\noindent
\[
\tilde{t}_i^2\leadsto \frac 1{\nu ^2+\sigma _u^2}.
\]

\vspace*{-0.5em}

\noindent
Therefore, from (\ref{eq:prf:lhs}), we obtain that

\noindent
\[
\frac 1{\sqrt{N}}[W^{-1}\!A^{\top }\!(b-\!P_nb)]_i\leadsto z_i\!\doteq
\!\left\{ \!
\begin{array}{l}
\vspace*{-6.0mm} \\
0,\hspace*{3mm}\mbox{for }i=1,\ldots ,n; \\
w_i^{-1}\nu \kappa h(i), \\
\hspace*{6mm}\mbox{for }i=n+1,\ldots ,q.\vspace*{-1.0mm}
\end{array}
\right.
\]

\vspace*{-0.5em}

\noindent where $\kappa \doteq \nu /\sqrt{\nu ^2+\sigma _u^2}$.

\noindent From the definition of the symbol $\leadsto $, the above
expression implies that for any given $\epsilon _1>0$ and $\beta _1\in (0,1)$
there exists an integer $N_1$ such that, for any $N\!\geq \!N_1$, it results

\noindent
\begin{equation}
\mathbb{P}\left\{ \left| \tfrac 1{\sqrt{N}}|[W^{-1}A^{\top
}\!(b-\!P_nb)]_i|\!-\!|z_i|\right| \geq \epsilon _1\!\right\} \leq \beta _1.
\label{eq:prob_beta_1}
\end{equation}

\vspace*{-0.5em}

\noindent Further, under the Assumption~\ref{ass:main_stable} that $%
|h(i)|\leq L\rho ^{i-1}$ and since the weight sequence is assumed to be
nondecreasing, we have that

\noindent
\begin{equation}
\hspace*{7mm}|z_i|\leq w_n^{-1}\nu \kappa L\rho ^n,\quad \forall i=1,\ldots
,q.  \label{lrho_ub}
\end{equation}

\vspace*{-0.5em}

\noindent Since, for all $i=1,\ldots ,q$,

\noindent
\[
\begin{array}{l}
\left| \frac 1{\sqrt{N}}|[W^{-1}A^{\top }(b-P_nb)]_i|-|z_i|\right| \geq \\
\hspace*{5mm}\geq \tfrac 1{\sqrt{N}}|[W^{-1}A^{\top }(b-P_nb)]_i|-|z_i| \\
\hspace*{5mm}\geq \tfrac 1{\sqrt{N}}|[W^{-1}A^{\top
}(b-P_nb)]_i|-w_n^{-1}\nu \kappa L\rho ^n,
\end{array}
\]

\vspace*{-0.5em}

\noindent from (\ref{eq:prob_beta_1}) it follows that, for any $N\!\geq
\!N_1 $,

\noindent
\[
\mathbb{P}\!\left\{ \!\tfrac 1{\sqrt{N}}|[W^{-1}A^{\top
}\!(b-\!P_nb)]_i|-w_n^{-1}\nu \kappa L\rho ^n\geq \epsilon _1\!\right\} \leq
\beta _1;\label{eq:prob_beta_2}
\]

\vspace*{-0.5em}

\noindent hence, from Bonferroni's inequality, for any $N\!\geq \!N_1$ we
have

\noindent
\[
\mathbb{P}\!\left\{ \!\tfrac 1{\sqrt{N}}\Vert W^{-1}A^{\top
}\!(b-\!P_nb)\Vert _\infty \!>w_n^{-1}\nu \kappa L\rho ^n\!\!+\!\epsilon
_1\!\right\} \!\leq \!q\beta _1.\label{eq:prob_beta_3}
\]

\vspace*{-0.5em}

\noindent Taking the complementary event, for any $N\geq N_1$ it results

\noindent
\begin{equation}
\mathbb{P}\!\left\{ \!\tfrac 1{\sqrt{N}}\Vert W^{-1}A^{\top
}\!(b-\!P_nb)\Vert _\infty \!\leq \!w_n^{-1}\nu \kappa L\rho
^n\!\!+\!\epsilon _1\!\right\} \!\geq \!1-q\beta _1.  \label{eq:prob_beta_4}
\end{equation}

\vspace*{-0.0em}

\noindent Similarly, from (\ref{eq:upsilon_n}) it follows that for any given
$\epsilon _2>0$ and $\beta _2\in (0,1)$ there exists an integer $N_2$ such
that

\noindent
\begin{equation}
\mathbb{P}\{|\Upsilon _n(A)\!-\!1|\!\leq \!\epsilon _2\}\!=\!\mathbb{P}%
\{1\!-\!\Upsilon _n(A)\!\leq \!\epsilon _2\}\!\geq \!1\!-\!\beta
_2,\;\forall N\!\geq \!N_2;  \label{eq:prob_ups_1}
\end{equation}

\vspace*{-0.0em}

\noindent thus

\noindent
\begin{equation}
\mathbb{P}\left\{ \!\tfrac \gamma {2\sqrt{N}}\left( 1\!-\!\epsilon _2\right)
\leq \tfrac \gamma {2\sqrt{N}}\Upsilon _n\!(A)\!\right\} \geq 1\!-\!\beta
_2,\;\forall N\!\geq \!N_2.  \label{eq:prob_ups_2}
\end{equation}

\vspace*{-0.5em}

\noindent Considering the joint events in (\ref{eq:prob_beta_4}) and (\ref
{eq:prob_ups_2}), we have from Bonferroni's inequality that\vspace*{-1.2em}

\noindent
\begin{eqnarray*}
&&\hspace*{3mm}\left\{ \tfrac 1{\sqrt{N}}\Vert W^{-1}A^{\top
}\!(b-\!P_nb)\Vert _\infty \leq w_n^{-1}\nu \kappa L\rho ^n+\epsilon
_1\right\} \;\cap \\
&&\;\hspace*{17mm}\left\{ \tfrac \gamma {2\sqrt{N}}\left( 1-\epsilon
_2\right) \leq \tfrac \gamma {2\sqrt{N}}\Upsilon _n(A)\right\}
\end{eqnarray*}

\vspace*{-1.2em}

\noindent holds with probability no smaller than $1-\beta $, for any $N\geq
N_\beta \doteq \max (N_1,N_2)$, with $\beta \doteq q\beta _1+\beta _2$.
Next, observe that if it holds that

\noindent
\begin{equation}
w_n^{-1}\nu \kappa L\rho ^n+\epsilon _1\leq \tfrac \gamma {2\sqrt{N}}\left(
1-\epsilon _2\right) ,  \label{eq:condgam1}
\end{equation}

\vspace*{-0.5em}

\noindent then we may conclude with confidence at least $1-\beta $ that

\noindent
\begin{equation}
\tfrac 1{\sqrt{N}}\Vert W^{-1}A^{\top }(b-P_nb)\Vert _\infty \leq \tfrac
\gamma {2\sqrt{N}}\Upsilon _n(A).  \label{eq:condgam2}
\end{equation}

\vspace*{-0.5em}

\noindent Suppose that condition (\ref{eq:gamma_lower}) holds, thus $\gamma
\!=\!2\mu w_n^{-1}\!L\rho ^n\nu \kappa \sqrt{N}$ for some $\mu >1$;
substituting this expression into (\ref{eq:condgam1}), we obtain the
condition

\noindent
\[
\epsilon _1+\mu w_n^{-1}L\rho ^n\nu \kappa \epsilon _2\leq \left( \mu
-1\right) w_n^{-1}L\rho ^n\nu \kappa .
\]

\vspace*{-0.5em}

\noindent Since $\mu >1$, and since $\epsilon _1$, $\epsilon _2$ can be
chosen arbitrarily, this condition is satisfied for a sufficiently small
choice of $\epsilon _1$, $\epsilon _2$. Therefore, condition (\ref
{eq:condgam1}) is satisfied, and hence (\ref{eq:condgam2}) is satisfied with
probability no smaller than $1-\beta $. The statement then follows from
Lemma~\ref{lem:recovery_1}. \qed

\subsection{Proof of Corollary~\ref{cor:main_lead}}

\label{sec:app_proof:cor_main_lead} We apply Theorem~\ref{thm:main} with $n$
being equal to the leading order $n_l(N)$ of the system. Since (\ref
{eq:leadingorder}) holds for $i=n_l(N)$, substituting this expression into (%
\ref{eq:gamma_lower}) we have the condition

\noindent
\[
\gamma \geq 2\mu w_{n_l(N)}^{-1}\rho \sigma _y\kappa ,
\]

\vspace*{-0.5em}

\noindent for some $\mu >1$, which is equivalent to (\ref
{eq:gamma_lower_const}). The claim then follows by applying Theorem~\ref
{thm:main}. \qed

\subsection{Proof of Corollary~\ref{cor:main_fir}}

\label{sec:app_proof:cor_main} We follow the same reasoning as in Section~%
\ref{sec:app_proof:thm_main} up to (\ref{eq:prf:lhs}). Then, we observe that
since $h$ is FIR of order $n$, then $h_{\downarrow n}$ is identically zero,
hence from (\ref{eq:prf:lhs}) it follows that

\noindent
\[
\tfrac 1{\sqrt{N}}W^{-1}A^{\top }(b-P_nb)\leadsto 0,
\]

\vspace*{-0.5em}

\noindent which means that for any given $\epsilon _1>0$ and $\beta _1\in
(0,1)$ there exists an integer $N_1$ such that

\noindent
\[
\mathbb{P}\left\{ \tfrac 1{\sqrt{N}}\Vert W^{-1}A^{\top }(b-P_nb)\Vert
_\infty \leq \epsilon _1\right\} \geq 1-q\beta _1,\quad \forall N\geq N_1.%
\label{eq:prob_beta_4_c}
\]

\vspace*{-0.5em}

\noindent Following a reasoning similar to the one in (\ref{eq:prob_ups_1}%
)--(\ref{eq:condgam2}), we claim that if

\noindent
\begin{equation}
\epsilon _1\leq \tfrac \gamma {2\sqrt{N}}\left( 1-\epsilon _2\right) ,
\label{eq:condgam1_c}
\end{equation}

\vspace*{-0.5em}

\noindent then we may conclude with confidence at least $1-\beta $ that

\noindent
\begin{equation}
\tfrac 1{\sqrt{N}}\Vert W^{-1}A^{\top }(b-P_nb)\Vert _\infty <\tfrac \gamma
{2\sqrt{N}}\Upsilon _n(A).  \label{eq:condgam2_c}
\end{equation}

\vspace*{-0.5em}

\noindent But, since $\gamma >0$, condition (\ref{eq:condgam1_c}) can always
be satisfied for some $\epsilon _1,\epsilon _2$, and hence (\ref
{eq:condgam2_c}) holds with probability at least $1-\beta $. The claim then
follows from Lemma~\ref{lem:recovery_1}. \qed

\TeXButton{normalsize}{\normalsize}

\bibliographystyle{plainnat}
\bibliography{mybiblio}

\begin{thebibliography}{23}
\providecommand{\natexlab}[1]{#1}
\providecommand{\url}[1]{\texttt{#1}}
\expandafter\ifx\csname urlstyle\endcsname\relax
  \providecommand{\doi}[1]{doi: #1}\else
  \providecommand{\doi}{doi: \begingroup \urlstyle{rm}\Url}\fi

\bibitem[Akaike(1974)]{Akai74}
H.~Akaike.
\newblock A new look at the statistical model identification.
\newblock \emph{IEEE Transactions on Automatic Control}, AC-19\penalty0
  (6):\penalty0 716--723, 1974.

\bibitem[Bishop(1995)]{Bishop:95}
C.M. Bishop.
\newblock Training with noise is equivalent to {T}ikhonov regularization.
\newblock \emph{J. Neural Computation}, 7\penalty0 (1):\penalty0 108--116,
  1995.

\bibitem[Chen et~al.(2012)Chen, Ohlsson, and Ljung]{ChOL12}
T.~Chen, H.~Ohlsson, and L.~Ljung.
\newblock On the estimation of transfer funtions, regularizations and
  {G}aussian processes -- revisited.
\newblock \emph{Automatica}, 48\penalty0 (8):\penalty0 1525--1535, 2012.

\bibitem[{De Mol} et~al.(2009){De Mol}, {De Vito}, and Rosasco]{DeDeRo:09}
C.~{De Mol}, E.~{De Vito}, and L.~Rosasco.
\newblock Elastic-net regularization in learning theory.
\newblock \emph{J. of Complexity}, 25\penalty0 (2):\penalty0 201--230, 2009.

\bibitem[Friedman et~al.(2010)Friedman, Hastie, and Tibshirani]{FrHaTi:10}
J.~Friedman, T.~Hastie, and R.~Tibshirani.
\newblock Regularization paths for generalized linear models via coordinate
  descent.
\newblock \emph{J. of Statistical Software}, 33\penalty0 (1):\penalty0 1--22,
  2010.

\bibitem[Hastie et~al.(2009)Hastie, Tibshirani, and Friedman]{HaTiFr:09}
T.~Hastie, R.~Tibshirani, and J.~Friedman.
\newblock \emph{The {E}lements of {S}tatistical {L}earning : {D}ata {M}ining,
  {I}nference, and {P}rediction}.
\newblock Springer Series in Statistics. Springer, New York, second edition,
  2009.

\bibitem[Koll{\'{a}}r(1994)]{Koll94}
I.~Koll{\'{a}}r.
\newblock \emph{Frequency Domain System Identification Toolbox User's Guide}.
\newblock The MathWorks, Inc., Natick, MA, 1994.

\bibitem[Koll{\'{a}}r et~al.(1994)Koll{\'{a}}r, Pintelon, and
  Schoukens]{KoPS94}
I.~Koll{\'{a}}r, R.~Pintelon, and J.~Schoukens.
\newblock Frequency domain system identification toolbox for {M}atlab: a
  complex application example.
\newblock In \emph{Proc. {of} IFAC SYSID'94}, pages 23--28, vol.~4, Copenhagen,
  Denmark, 1994.

\bibitem[Ljung(1999{\natexlab{a}})]{Ljung:99_book}
L.~Ljung.
\newblock \emph{System Identification: {T}heory for the User}.
\newblock Prentice-Hall, Englewood Cliffs, second ed., 1999{\natexlab{a}}.

\bibitem[Ljung(1999{\natexlab{b}})]{Ljung:99_proc}
L.~Ljung.
\newblock Model validation and model error modeling.
\newblock In B.~Wittenmark and A.~Rantzer, editors, \emph{The {\AA}str\"om
  Symposium on Control}, pages 15--42, Lund, Sweden, Aug. 1999{\natexlab{b}}.
  Studentlitteratur.

\bibitem[Milanese et~al.(2010)Milanese, Ruiz, and Taragna]{MiRT10}
M.~Milanese, F.~Ruiz, and M.~Taragna.
\newblock Direct data-driven filter design for uncertain {LTI} systems with
  bounded noise.
\newblock \emph{Automatica}, 46\penalty0 (11):\penalty0 1773--1784, 2010.

\bibitem[{National Instruments Corporation}(2004-2006)]{LabVIEW06}
{National Instruments Corporation}.
\newblock \emph{{LabVIEW} System Identification Toolkit User Manual}.
\newblock Austin, TX, 2004-2006.

\bibitem[Pillonetto and {De Nicolao}(2010)]{PiDN10}
G.~Pillonetto and G.~{De Nicolao}.
\newblock A new kernel-based approach for linear system identification.
\newblock \emph{Automatica}, 46\penalty0 (1):\penalty0 81--93, 2010.

\bibitem[Pillonetto et~al.(2011)Pillonetto, Chiuso, and {De Nicolao}]{PiCD11}
G.~Pillonetto, A.~Chiuso, and G.~{De Nicolao}.
\newblock Prediction error identification of linear systems: a nonparametric
  {G}aussian regression approach.
\newblock \emph{Automatica}, 47\penalty0 (2):\penalty0 291--305, 2011.

\bibitem[Pillonetto et~al.(2014)Pillonetto, Dinuzzo, Chen, {De Nicolao}, and
  Ljung]{PDCD14}
G.~Pillonetto, F.~Dinuzzo, T.~Chen, G.~{De Nicolao}, and L.~Ljung.
\newblock Kernel methods in system identification, machine learning and
  function estimation: A survey.
\newblock \emph{Automatica}, 50\penalty0 (3):\penalty0 657--682, 2014.

\bibitem[Pintelon and Schoukens(2012)]{PiSc12}
R.~Pintelon and J.~Schoukens.
\newblock \emph{System identification: a frequency domain approach}.
\newblock John Wiley \& Sons, second edition, 2012.

\bibitem[Rissanen(1978)]{Riss78}
J.~Rissanen.
\newblock Modelling by shortest data description.
\newblock \emph{Automatica}, 14\penalty0 (5):\penalty0 465--471, 1978.

\bibitem[Schwarz(1978)]{Schw78}
G.~Schwarz.
\newblock Estimating the dimension of a model.
\newblock \emph{The Annals of Statistics}, 6\penalty0 (2):\penalty0 461--464,
  1978.

\bibitem[S{\"{o}}derstr{\"{o}}m and Stoika(1989)]{SoSt89}
T.~S{\"{o}}derstr{\"{o}}m and P.~Stoika.
\newblock \emph{System Identification}.
\newblock Prentice-Hall, 1989.

\bibitem[Tibshirani(1996)]{Tibshirani:96}
R.~Tibshirani.
\newblock Regression shrinkage and selection via the lasso.
\newblock \emph{J. Royal Statist. Soc. B}, 58\penalty0 (1):\penalty0 267--288,
  1996.

\bibitem[Torfs et~al.(1998)Torfs, Vuerinckx, Swevers, and Schoukens]{TVSS98}
D.~E. Torfs, R.~Vuerinckx, J.~Swevers, and J.~Schoukens.
\newblock Comparison of two feedforward design methods aiming at accurate
  trajectory tracking of the end point of a flexible robot arm.
\newblock \emph{IEEE Transactions on Control Systems Technology}, 6\penalty0
  (1):\penalty0 2--14, January 1998.

\bibitem[Tropp(2006)]{Tropp:06}
J.~A. Tropp.
\newblock Just relax: convex programming methods for identifying sparse signals
  in noise.
\newblock \emph{{IEEE} Transactions on Information Theory}, 52\penalty0
  (3):\penalty0 1030--1051, 2006.

\bibitem[Zou and Hastie(2005)]{ZouHas:05}
H.~Zou and T.~Hastie.
\newblock Regularization and variable selection via the elastic net.
\newblock \emph{J. Royal Statist. Soc. B}, 67\penalty0 (2):\penalty0 301--320,
  2005.

\end{thebibliography}

\end{document}